\documentclass[11pt]{article}
\pdfoutput=1
\usepackage[utf8]{inputenc}
\usepackage{dsfont}
\usepackage{diagbox}
\usepackage{amsfonts}
\usepackage{mathtools}
\allowdisplaybreaks[4]        
\usepackage{amssymb}
\usepackage{euscript}     
\usepackage{braket}
\usepackage{starfont}
\usepackage{color,soul}         
\usepackage{braket}
\usepackage{tensor}        
\usepackage{amsthm}
\usepackage{graphicx}
\usepackage{slashed}
\usepackage{leftidx}
\usepackage{subfigure}
\usepackage{bbm}
\definecolor{outerspace}{rgb}{0.25, 0.29, 0.3}
\definecolor{scarlet}{rgb}{1.0, 0.13, 0.0}
\usepackage[header,title,page,titletoc]{appendix}  
\definecolor{princetonorange}{rgb}{1.0, 0.56, 0.0}
\definecolor{WildStrawberry}{rgb}{1.0, 0.26, 0.64}
\definecolor{rossocorsa}{rgb}{0.83, 0.0, 0.0}
\definecolor{navyblue}{rgb}{0.0, 0.0, 0.5}
\usepackage[numbers,sort&compress]{natbib}  
\usepackage{float}
\usepackage[paper=letterpaper,margin=1in]{geometry}
\newcommand{\req}[1]{(\ref{#1})} %{Eq.\thinspace(\ref{#1})}  

\newcommand{\bea}{\begin{eqnarray}}
\newcommand{\diff}{\mathrm{d}}
\newcommand{\eea}{\end{eqnarray}}
\newcommand{\ba}{\begin{eqnarray}}
\newcommand{\ea}{\end{eqnarray}}

\newcommand{\be}{\begin{equation}}
\newcommand{\ee}{\end{equation} }
\newcommand{\beqa}{\begin{eqnarray}}
\newcommand{\eeqa}{\end{eqnarray}}
\newcommand{\beqar}{\begin{eqnarray*}}
\newcommand{\eeqar}{\end{eqnarray*}}

\renewcommand{\req}[1]{eq.~(\ref{#1})}

\newcommand{\ssc}{\scriptscriptstyle}
\newcommand{\eg}{{\it e.g.,}\ }
\newcommand{\ie}{{\it i.e.,}\ }

\newcommand{\see}{S_{\ssc \rm EE}}

\DeclareMathOperator{\tr}{tr}

\usepackage{hyperref}
\hypersetup{
    colorlinks,
    citecolor=rossocorsa,
    filecolor=navyblue,
    linkcolor=navyblue,
    urlcolor=navyblue
}

\begin{document} 

\begin{titlepage}

\begin{center}

\phantom{ }
\vspace{3cm}

{\bf \Large{Reflected entropy for free scalars}}
\vskip 0.5cm
Pablo Bueno${}^{\text{\Zeus}}$ and Horacio Casini${}^{\text{\Kronos}}$
\vskip 0.05in
\textit{Instituto Balseiro, Centro At\'omico Bariloche}
\vskip -.4cm
\textit{ 8400-S.C. de Bariloche, R\'io Negro, Argentina}

%\vskip -.4cm
%\small{\textit{ C/ Nicol\'as Cabrera, 13-15, C.U. Cantoblanco, 28049 Madrid, Spain}}

%\small{${}^{\text{\Apollon}}$ \textit{Department of Mathematics and Statistics, Memorial University of Newfoundland}}
%\vskip -.4cm
%\small{\textit{  St. John's, Newfoundland and Labrador, A1C 5S7, Canada}}

%\vskip -.4cm
%\small{\textit{}}
\begin{abstract}
%Recently, a novel notion ... This Type-I or ``reflected'' entropy
%We continue our study of reflected entropy, $R(A,B)$, for Gaussian systems. In this paper we provide general formulas valid for free scalar fields in arbitrary dimensions. Similarly to the fermionic case, the resulting expressions are fully determined in terms of correlators of the fields, making them amenable to lattice calculations. We apply this to the case of a $(1+1)$-dimensional chiral scalar, whose reflected entropy we compute for two intervals as a function of the cross-ratio, comparing it with previous holographic and free-fermion results. For both types of free theories we find that reflected entropy satisfies the conjectural monotonicity property $R(A,BC) \geq R(A,B)$. Then, we move to $(2+1)$ dimensions and evaluate it for two length-$L$ squares separated a distance $\ell$ both for free scalars and fermions. From this, we compute the coefficient $\kappa^{(R)}$ controlling the linear behavior $R(A,B) \simeq \kappa^{(R)} x$ as $x\equiv L/\ell \gg 1$ for the free scalar, the free fermion and holographic Einstein gravity, and compare it to the analogous mutual information ones. In all cases considered, both for $(1+1)$- and $(2+1)$-dimensional theories, we verify that the general inequality relating both quantities, $R(A,B)\geq I(A,B)$, is satisfied. Based on our results we conjecture that, for general regions, their quotient behaves as $R(x)/I(x) \sim -\log x$ in the $x\ll 1$ regime (very far from each other) for general CFTs in arbitrary dimensions. 

We continue our study of reflected entropy, $R(A,B)$, for Gaussian systems. In this paper we provide general formulas valid for free scalar fields in arbitrary dimensions. Similarly to the fermionic case, the resulting expressions are fully determined in terms of correlators of the fields, making them amenable to lattice calculations. We apply this to the case of a $(1+1)$-dimensional chiral scalar, whose reflected entropy we compute for two intervals as a function of the cross-ratio, comparing it with previous holographic and free-fermion results. For both types of free theories we find that reflected entropy satisfies the conjectural monotonicity property $R(A,BC) \geq R(A,B)$. Then, we move to $(2+1)$ dimensions and evaluate it for square regions for free scalars, fermions and holography, determining the very-far and very-close regimes and comparing them with their mutual information counterparts.
%From this, we compute the coefficient $\kappa^{(R)}$ controlling the linear behavior $R(A,B) \simeq \kappa^{(R)} x$ as $x\equiv L/\ell \gg 1$ for the free scalar, the free fermion and holographic Einstein gravity, and compare it to the analogous mutual information ones. 
In all cases considered, both for $(1+1)$- and $(2+1)$-dimensional theories, we verify that the general inequality relating both quantities, $R(A,B)\geq I(A,B)$, is satisfied. Our results suggest that for general regions characterized by length-scales $L_A\sim L_B\sim L$ and separated a distance $\ell$, the reflected entropy in the large-separation regime ($x\equiv L/\ell \ll 1$) behaves as $R(x) \sim - I(x) \log x$ for general CFTs in arbitrary dimensions.

\end{abstract}
\end{center}

\small{\vspace{5cm}\noindent${}^{\text{\text{\Zeus}}}$pablo.bueno$@$cab.cnea.gov.ar\\
${}^{\text{\Kronos}}$casini@cab.cnea.gov.ar}

\end{titlepage}

\setcounter{tocdepth}{2}

{\parskip = .2\baselineskip \tableofcontents}

\section{Introduction}
Entanglement entropy (EE) of subregions is an ill-defined quantity in quantum field theory (QFT). This fact can be understood from various perspectives. From a lattice point of view, as we reduce the lattice spacing a growing amount of entanglement across the entangling surface adds up, producing the usual area-law divergence (and others) in the limit. From the continuum theory perspective, the underlying reason has to do with the fact that algebras of operators associated to spatial regions are von Neumann algebras of type-III, for which all traces are either vanishing or infinite ---see \eg \cite{haag,Witten:2018lha}.

The situation improves when one considers two (or more) disjoint regions: entanglement measures such as mutual information $I(A,B)$ do make sense in QFT. The whole issue with the type-III-ness of subregion algebras has to do with the sharp spatial cut introduced by the entangling surface $\partial A$. When instead of considering a region and its complement, we consider two disjoint regions $A,B$, the so-called ``split-property''\footnote{This property holds in general under very mild assumptions related to the growth of the number of degrees of freedom at high energies, \cite{Buchholz:1973bk,Buchholz:1986dy}.} guarantees the existence of a tensor product decomposition of the global Hilbert space as $\mathcal{H}=\mathcal{H}_{\mathcal{N}_{AB}}\otimes \mathcal{H}_{\mathcal{N}_{AB}'}$ where $\mathcal{N}_{AB}$ and its commutant $\mathcal{N}'_{AB}$
 are type-I factors. The idea is that there always exists one such factor $\mathcal{N}_{AB}$ which contains the algebra of the first region, $\mathcal{A}_A$, while still commuting with the operators in algebra of the second, $\mathcal{A}_B$. Namely, one has $\mathcal{A}_A \subseteq \mathcal{N}_{AB} \subseteq (\mathcal{A}_B)'$. Importantly, contrary to $\mathcal{A}_A$ or $\mathcal{A}_B$, $\mathcal{N}_{AB}$ cannot be sharply associated to any particular geometric region\footnote{See our previous paper \cite{Bueno:2020vnx} for a possible notion of spatial ``algebra density'' in the case of free fermions.}. There is no problem in defining traces for Type-I von Neumann algebras and so given $\mathcal{N}_{AB}$, we can define the corresponding von Neumann entropy $S(\mathcal{N}_{AB})$ as the entropy of the reduced state in  any of the factors of the tensor product. 
 
 Now, there are infinitely many possible splits associated to a pair of regions $A,B$, so which one to choose? Interestingly, given a state which is cyclic and separating for the various algebras (\eg the vacuum), there is a somewhat canonical choice. This is \cite{Doplicher:1982cv,Doplicher:1984zz,Doplicher:1983if}
 \be
\mathcal{N}_{AB}\equiv \mathcal{A}_A \vee J_{AB}   \mathcal{A}_A J_{AB}\,,\quad \text{with the commutant given by} \quad 
\mathcal{N}_{AB}'=\mathcal{A}_B \vee J_{AB}   \mathcal{A}_B J_{AB}\, . \label{tomo}
\ee
Here we used the standard notation $\mathcal{A} \vee \mathcal{B}$ to refer to the double commutant of the algebra of the union, namely, $\mathcal{A} \vee \mathcal{B}\equiv (\mathcal{A} \cup \mathcal{B})''$.
Also, $J_{AB}$ is the Tomita-Takesaki modular conjugation operator associated to the algebra of $AB$ and the corresponding state. The von Neumann entropy associated to this type-I factor defines the reflected entropy \cite{Longo:2019pjj}
\begin{equation}\label{rsss}
R(A,B) \equiv S(\mathcal{N}_{AB})\, .
\end{equation}
An alternative route to the same notion was presented by Dutta and Faulkner in \cite{Dutta:2019gen}. A given state $\rho_{AB} $ in a Hilbert space $\mathcal{H}_A \otimes \mathcal{H}_B$ can be canonically purified as $\ket{ \sqrt{\rho_{AB}}}\in (\mathcal{H}_A \otimes \mathcal{H}_A^*) \otimes (\mathcal{H}_B \otimes \mathcal{H}_B^*)$. Then, the von Neumann entropy associated to the reduced density matrix $\rho_{AA^*}$ obtained from tracing out over $\mathcal{H}_B \otimes \mathcal{H}_B^*$ is nothing but the reflected entropy. Indeed, the modular conjugation operator $J_{AB}$ precisely maps $\mathcal{A}_A$ into $\mathcal{A}_{A^*}$, and one has $\mathcal{N}_{AB}=\mathcal{A}_{AA^*}$.
While this construction is not directly suitable for QFTs, one can safely use it in the lattice and unambiguously recover reflected entropy as defined in \req{rsss} in the continuum limit. A useful construction in terms of replica-manifold partition functions was also presented in that paper. In addition, they also showed that reflected entropy generally bounds above the mutual information. Namely,
\begin{equation}\label{ir}
R(A,B)\geq I(A,B)\, ,
\end{equation}
holds for general theories. 

Much of the interest in reflected entropy so far has come from the observation, by the same authors, that for holographic theories dual to Einstein gravity, this quantity is proportional to the minimal entanglement wedge cross section, $R_{\rm holo.}(A,B)= 2 E_{W}(A,B)$, at leading in order in Newton's constant \cite{Dutta:2019gen}. Subsequent work studying aspects of reflected entropy building up on the results of  \cite{Dutta:2019gen} includes \cite{Jeong:2019xdr,Kusuki:2019rbk,Akers:2019gcv,Kusuki:2019evw,Moosa:2020vcs,Kudler-Flam:2020url,Boruch:2020wbe,Asrat:2020uib,Kudler-Flam:2020yml,BabaeiVelni:2020wfl,Nakata:2020fjg,Li:2020ceg,Chandrasekaran:2020qtn}.  Candidates for multipartite versions of reflected entropy have also been proposed in \cite{Bao:2019zqc,Chu:2019etd,Marolf:2019zoo}. In passing, let us mention that $E_{W}$ has also been proposed to be related to the  ``entanglement of purification'' \cite{Takayanagi:2017knl,Nguyen:2017yqw} and to the so-called ``odd entropy'' \cite{Tamaoka:2018ned}. Regarding the latter, a similar connection between reflected entropy and odd entropy has been observed in \cite{Berthiere:2020ihq} in the case of Chern Simons theories in $(2+1)$ dimensions, although it is expected that both quantities differ in general \cite{Dutta:2019gen}. 

So far, it has not been rigorously proven that reflected entropy should be finite in general\footnote{Except when $A,B$ stop being disjoint. In fact, reflected entropy can be used as a geometric regulator for entanglement entropy \cite{Dutta:2019gen}, similarly to mutual information \cite{Casini:2006ws,Casini:2015woa}.}, although it is believed to be so at least for most QFTs ---see \cite{Longo:2019pjj} and also \cite{Narnhofer:2002ic,Otani:2017pmn,Hollands:2017dov}. This was proven to be the case for free fermions in $(1+1)$ dimensions in  \cite{Longo:2019pjj} and confirmed later in \cite{Bueno:2020vnx}, where we explicitly evaluated it for that theory as a function of the conformal cross ratio. The calculations in \cite{Berthiere:2020ihq} also yield finite answers.

The main purpose of this paper is to continue developing the general technology required for the evaluation of reflected entropy for Gaussian systems. As mentioned above, this was started in our previous paper \cite{Bueno:2020vnx}, where we obtained general formulas valid for free fermions in arbitrary dimensions. The focus here will be on free scalars, for which we will provide analogous expressions. This is the subject of section \ref{refffsc}. Analogously to the fermions case, we show that reflected entropy can be computed in terms of correlators of the bosonic fields associated to the system $A$. General formulas valid in general dimensions are presented both in the case in which the system is described in terms of $N$ scalars and $N$ conjugate momenta as well as in the case corresponding to a unified description in terms of $2N$ Hermitian operators. The main formulas are eqs. (\ref{phiii}), (\ref{g}) and (\ref{entroo}) in the first case and eqs. (\ref{gammaa3}), (\ref{F113}), (\ref{vy}) and (\ref{svn}) in the second.

We apply this formulas to the case of a chiral scalar in $(1+1)$ dimensions in section \ref{chiralsca}. We compute reflected entropy for this model for a pair of intervals as a function of the conformal cross ratio, and compare the result (normalized by the central charge) with the holographic \cite{Dutta:2019gen} and fermionic ones \cite{Bueno:2020vnx}. The scalar curve turns out to be considerably lower than the other two, but still greater than the mutual information in the whole range, as expected by the general inequality \req{ir}. In this section we also study how the type-I character of the algebra $\mathcal{N}_{AB}$ manifests itself in the structure of eigenvalues of the matrix of correlators required for the evaluation of reflected entropy as compared to the entanglement entropy one. As opposed to the latter, in the case of reflected entropy only a few eigenvalues make a relevant contribution to the result in the continuum. In this section we also verify the conjectured monotonicity of reflected entropy under inclusions for scalars and fermions. 

In section \ref{2plus1} we start the study of reflected entropy for $(2+1)$-dimensional free theories. In particular, we evaluate $R(A,B)$ for free scalars and fermions for regions $A,B$ corresponding to pairs of parallel squares of length $L$ separated a distance $\ell$. In both cases we find a finite answer as a function of $x\equiv L/\ell$ and verify that \req{ir} holds. Also, we observe that reflected entropy behaves linearly with $x$ as this quotient grows, $R(A,B) \simeq \kappa^{(R)} x$, analogously to mutual information. We compute the coefficient $\kappa^{(R)}$ numerically for both theories as well as for holography (using the connection with  $E_W$) and compare it to the respective mutual information answers. In the opposite  regime, \ie for $ x \ll 1$, we observe that $R(x) \sim - I(x) \log x$ holds for both free theories. The same behavior is found to occur for the $(1+1)$-dimensional theories considered in section \ref{chiralsca}, which leads us to conjecture that this is a general relation valid for arbitrary regions far apart from each other in general $d$-dimensional CFTs. 

We conclude with some future directions in section \ref{finnnn}. Appendix \ref{ape} contains a table with the numerical results found for the reflected entropy of $(1+1)$-dimensional free scalars and fermions for various values of the cross ratio.

\section{Reflected entropy for free scalars}\label{refffsc}
In this section we show how the reflected entropy for Gaussian scalar systems in general dimensions can be computed ---analogously to the entanglement entropy, and similarly to the fermion case explored in \cite{Bueno:2020vnx}--- from matrices of two-point functions of the scalar and conjugate-momentum fields. We also discuss how this formula gets modified when the usual description in terms of a set of scalars and momenta $\{\phi_i,\pi_j \}$, $i,j=1,\dots,N$, is replaced by one in terms of $2N$ Hermitian operators $f_i$, $i=1,\dots,2N$, more suitable in certain cases, such as the one corresponding to a chiral scalar in $d=2$.

\subsection{Purification and general formulas: take one}\label{tone}
%\comment{copy pasted from reflected fermions:}
Let us start with some general comments about purifications and Tomita-Takesaki theory ---see \eg \cite{Borchers:2000pv} for a review of the latter.
Consider a quantum mechanical system with Hilbert space  ${\cal H}_1$ and an invertible density matrix $\rho$ written in its spectral decomposition
%Let $\rho$ be an invertible density matrix in a general quantum mechanical system of Hilbert space ${\cal H}_1$.  We can write 
 \begin{equation} \label{rori}
 \rho=\sum_p \lambda_p \vert p\rangle\langle p\vert\, .
 \end{equation}
%  where $\lambda_p$ is the eigenvalue of $\rho$ corresponding to the eigenvector $\vert p\rangle$.
Let us now consider a copy of ${\cal H}_1$, which we denote by ${\cal H}_2$. We can define a purification $\vert \Omega \rangle$ of $\rho$ in ${\cal H}_1\otimes{\cal H}_2$, so that $\rho=\textrm{tr}_{{\cal H}_2}\vert \Omega\rangle\langle \Omega\vert$.  In the Schmidt basis, this can be written as
\begin{equation}
\vert \Omega \rangle = \sum_p \sqrt{\lambda_p}  \vert p \,\tilde{p}\rangle\,.\label{laba}
\end{equation}
Observe that the orthonormal basis $\{\vert\tilde{p}\rangle\}$ for ${\cal H}_2$ in (\ref{laba}) is arbitrary, different choices corresponding to different purifications of $\vert \Omega \rangle$. As far as reflected entropy is concerned, all these choices are equivalent.  
 
Modular conjugation $J$ is defined by the anti-unitary operator 
\begin{equation}
J\equiv \sum_{p q} \vert p \,\tilde{q}\rangle \langle q\,\tilde{p}\vert \, *\,,\label{jota}
\end{equation}
where $*$ denotes complex conjugation in the basis $\{\vert p \tilde{q} \rangle\}$. 
One has  $J\vert \Omega\rangle=\vert \Omega\rangle$, $J^2=1$, $J^{\dagger}=J^{-1}=J$. Another important property is that the conjugation of an operator acting on the first factor produces an operator acting on the second, 
\begin{equation}
J({\cal O}\otimes 1)J=1\otimes \bar{{\cal O}}\,.
\end{equation}
Now, defining 
$
\Delta\equiv \rho\otimes \rho^{-1}\,,
$
the Tomita-Takesaki relations follow,
\be \label{ttr}
J\, \Delta=\Delta^{-1} \,J\,,\hspace{1cm} J \Delta^{1/2} {\cal O}_1 | \Omega \rangle={\cal O}_1^{\dagger} | \Omega \rangle \,,
\ee
where ${\cal O}_1$ is any operator acting on the first factor.

Let us now focus our discussion on free scalar fields.
Let $\phi_i$, and $\pi_j$, $i,j=1,...,N$, be a system of scalars and conjugate momenta acting on a Hilbert space ${\cal H}_1$. These are Hermitian operators which satisfy canonical commutation relations
\begin{equation}\label{comuu}
[\phi_i,\pi_j]=i\delta_{ij}\, , \quad [\phi_i,\phi_j]=[\pi_i,\pi_j]=0\, .
\end{equation}
Given a density matrix  $\rho\in \mathcal{H}_1$, we can purify it by considering a Hilbert space ${\cal H}$ of double dimension and extend the bosonic algebra with $2N$ additional operators $\phi_i,\pi_j$ so that \req{comuu} holds for $i,j=1,\dots,2N$. This can be achieved by defining
\begin{equation}\label{till}
\tilde{\phi}_i\equiv J \phi_i J \, , \quad \tilde{\pi}_j\equiv -J\pi_j J\, .
\end{equation}
Then it follows that the set $\{(\phi_1,\pi_1),\dots, (\phi_N, \pi_N),(\tilde \phi_1,\tilde \pi_1),\dots, (\tilde \phi_N,\tilde \pi _N) \}$ forms a canonical algebra of Hermitian operators in the full space ---in particular, \req{comuu} holds for all variables.

With these definitions, scalar correlators depend only on the density matrix $\rho$ for the first $N$ scalars. In order to see this, let us define $\Psi_i^0\equiv \phi_i $, $\Psi_i^1\equiv \pi_i$, and the same for $\tilde \Psi^a_i$, $a=0,1$. We have, in the purified state $\ket{\Omega}$ in the full space, 
\begin{align}\label{pspsp}
\braket{\Omega|\Psi_{i_1}^{a_1}\cdots \Psi_{i_k}^{a_k} \tilde \Psi_{j_1}^{b_1}\cdots  \tilde \Psi_{j_l}^{b_l}|\Omega}&=(-1)^{\sum_l b_l}\braket{\Omega|\Psi_{i_1}^{a_1}\cdots \Psi_{i_k}^{a_k} J  \Psi_{j_1}^{b_1}\cdots  \Psi_{j_l}^{b_l}|\Omega}\\ \notag
&=(-1)^{\sum_l b_l}\braket{\Omega|\Psi_{i_1}^{a_1}\cdots \Psi_{i_k}^{a_k} \Delta^{1/2}  \Psi_{j_l}^{b_l}\cdots  \Psi_{j_1}^{b_1}|\Omega}\\ \notag &=(-1)^{\sum_l b_l} \tr \left(\rho^{1/2} \Psi_{i_1}^{a_1}\cdots \Psi_{i_k}^{a_k} \rho^{1/2}  \Psi_{j_l}^{b_l}\cdots  \Psi_{j_1}^{b_1} \right)\, .
\end{align}
The first equal follows from \req{till} and the properties of the modular conjugation. The second, from \req{ttr} and the Hermiticity of the fields. The third can be easily verified using \req{rori} and \req{laba} explicitly.

Now, consider a set of creation and annihilation operators $a_l,a_l^{\dagger}$, $l=1,\dots,N$, satisfying $[a_i,a_j^{\dagger}]=\delta_{ij}$, related to the $\phi_i$ and $\pi_j$ via linear combinations
\begin{equation}
\phi_i = \alpha_{ij}\left[ a^{\dagger}_j +a_j\right] \, , \quad \pi_i=i \beta_{ij} \left[ a_j-a_j^{\dagger}\right]\, ,
\end{equation}
where $\alpha$ and $\beta$ are real matrices \cite{Casini:2009sr}.
The commutation relations in \req{comuu} impose the constraint $\alpha=-\frac{1}{2} (\beta^{T})^{-1}$.

The idea is now to assume a density matrix $\rho$ of the form \cite{2003JPhA...36L.205P,Chung_2000}
\begin{equation}
\rho= \Pi_l (1-e^{-\epsilon_l}) e^{-\sum_l \epsilon_l a_l^{\dagger} a_l}\, ,
\end{equation}
which defines a Gaussian state.
The two-point correlators of the fields and momenta will be denoted by (this notation is somewhat standard for correlators in general states)
\begin{equation}
X_{ij}\equiv \tr(\rho \phi_i \phi_j)\, , \quad P_{ij} \equiv  \tr(\rho \pi_i \pi_j)\, .
\end{equation}
On the other hand, for Gaussian states invariant under time reflection, we have \cite{Casini:2009sr}
\begin{equation}
\tr(\rho \phi_i \pi_j)=\tr(\rho \phi_i \pi_j)^* = \frac{i}{2} \delta_{ij}\, .
\end{equation}
These matrices of correlators can be written in terms of the expectation value of the number operator $n_{kk}\equiv \braket{a_k^{\dagger}a_k}=(e^{\epsilon_k}-1)^{-1}$. The results read
%\begin{equation}
%\alpha^* n \beta^T - \alpha (n+1) \beta^{\dagger}=\frac{1}{2}\, , \quad \alpha^* n \alpha^T + \alpha (n+1) \alpha^{\dagger}%=X\, , \quad  \beta^* n \beta^T + \beta (n+1) \beta^{\dagger}=P\, .
%\end{equation}
\begin{equation}\label{XPn}
\alpha( 2 n+1) \alpha^T =X\, , \quad \frac{1}{4} (\alpha^{-1})^T (2n+1) (\alpha^{-1})=P \quad \Rightarrow \quad  \frac{1}{4}\alpha (2n+1)^2 \alpha^{-1}= XP \, .
\end{equation}
 %Using this, it is possible to relate the spectrum of $\rho$ with the one of $C\equiv \sqrt{XP}$. One has
% \begin{equation}
% \nu_k =\frac{1}{2} \coth (\epsilon_k/2)\, ,
 %\end{equation}
 %where $\nu_k$ are the eigenvalues of $C$.
Going back to our double Hilbert space, the purified state $\ket{\Omega}$ is also Gaussian for the full system of scalars.%\comment{more comments on this?}

Organizing the scalars in a single field $\Phi_i \equiv \phi_i$, $i=1,\dots,N$ and $\Phi_{i+N}\equiv \tilde \phi_i$, $i=1,\dots,N$, and proceeding similarly for the momenta, $\Pi_i \equiv \pi_i$, $i=1,\dots,N$ and $\Pi_{i+N}\equiv \tilde \pi_i$, $i=1,\dots,N$, we are interested in the following correlators
\begin{equation}
\Phi_{ij}\equiv \braket{ \Omega |  \Phi_i \Phi_j | \Omega } \, , \quad \Pi_{ij} \equiv \braket{ \Omega |  \Pi_i \Pi_j | \Omega } \, , \quad i=1,\dots,2N\, .
\end{equation}
 Using \req{pspsp} we obtain the following block-matrix representation of these two objects
  \begin{align} \label{phi1}
\Phi&=\left(
\begin{array}{cc}
\alpha (2n+1) \alpha ^T & 2 \alpha \sqrt{n(n+1)} \alpha^T\\
2 \alpha \sqrt{n(n+1)} \alpha^T & \alpha (2n+1) \alpha^T
\end{array}
\right)\, , \\     \Pi&=\left(
\begin{array}{cc}
\frac{1}{4} (\alpha^{-1})^T (2n+1)\alpha^{-1} &-\frac{1}{2} (\alpha^{-1})^T \sqrt{n(n+1)} \alpha^{-1}\\
-\frac{1}{2} (\alpha^{-1})^T \sqrt{n(n+1)} \alpha^{-1} & \frac{1}{4} (\alpha^{-1})^T (2n+1)\alpha^{-1}
\end{array}
\right)\, .  \label{pi1}
\end{align}
These can be written in terms of $X$ and $P$ alone as
  \begin{align} \label{phiii}
\Phi=\left(
\begin{array}{cc}
X &  g(XP) X\\
g(XP) X & X
\end{array}
\right)\, , \quad    \Pi=\left(
\begin{array}{cc}
P &-  Pg(XP)  \\
- Pg(XP)   & P
\end{array}
\right)\, ,
\end{align}
where
\begin{equation}\label{g}
g(A)\equiv \sqrt{A-1/4}\sqrt{A}^{-1}\, .
\end{equation}
 %blah
 % \begin{equation}
%n=\sqrt{\alpha^{-1} XP \alpha}-\frac{1}{2} 
% \end{equation}
% and \req{XPn}.
 %\begin{equation}
%\Phi=\left(
%\begin{array}{cc}
%X & 2 \sqrt{X P -1/4}\\
%2 \sqrt{X P -1/4} & X
%\end{array}
%\right)\, , \quad      \Pi=\left(
%\begin{array}{cc}
%P & -\frac{1}{2} \sqrt{X P -1/4}\\
 %-\frac{1}{2} \sqrt{X P -1/4} & P%
%\end{array}
%\right)\, .  \label{xij}
%\end{equation}
%\comment{Here I used $f(A)g(A)=g(A)f(A)$ for $A=PX$ and $f(x)=\sqrt{x}^{-1}$, $g(x)=\sqrt{x-1/4}$}\\
The purity of the global state imposes that these matrices satisfy $\Phi \Pi=1/4$,  which can be easily verified. %as can be verified using the relation $P=X/4$, which follows from setting $\alpha=1$ above. 

Now, the von Neumann entropy corresponding to a region $Y$ can be obtained from the restriction of $\Phi$ and $\Pi$ to $Y$, \ie $(\Phi_Y)_{ij}=\Phi_{ij}$ and $(\Pi_Y)_{ij}=\Pi_{ij}$ for all $i,j\in Y$. Defining $C_{Y}\equiv \sqrt{\Phi_Y \Pi_Y}$, the entropy is given by 
 \begin{equation}\label{entroo}
 S(Y)=\tr \left[(C_Y+1/2) \log (C_Y+1/2) - (C_Y-1/2)\log (C_Y-1/2) \right]\, .
 \end{equation}
 In the continuum, the same expression can be used, where $C_Y$ is to be understood as a kernel, $C(x,y)$, $x,y\in Y$.
 
 When computing reflected entropy for a pair of regions $A$, $B$, we need to evaluate the $X$, $P$ and $g(XP)$ matrices for all sites belonging to those regions, which allows us to build the $\Phi$ and $\Pi$ matrices, and then restrict the different blocks to the region $A$ sites  ---see below for explicit examples. Formulas \req{entroo} and \req{phiii} can be thought of as generalizations of the well-known expressions required for the evaluation of the usual entanglement entropy ---see \eg \cite{Casini:2009sr}. In that case, \req{entroo} holds, where the matrix $C_Y$ is now the restriction to the entangling region $Y$ of the matrix $C_Y\equiv \sqrt{X_Y P_Y}$. In the reflected entropy case, \req{entroo} computes the entropy for $\rho_{AA^*}$ instead of $\rho_A$. The difference between both cases is codified in the additional blocks appearing in $\Phi$ and $\Pi$ with respect to $X$ and $P$ respectively.

 \subsection{Purification and general formulas: take two}\label{take2}
 The previous description in terms of scalar and conjugate-momentum fields can be generalized by considering instead a set of $2N$ Hermitian operators $f_i$ satisfying commutation relations of the form
 \begin{equation}
 [f_i,f_j]=i \left( \delta_{j i+1}-\delta_{j i-1}\right)\equiv i C_{ij}\, .
 \end{equation}
This is a more suitable choice in some cases, such as the one corresponding to a $d=2$ chiral scalar, which we consider in the following section. This setup has been previously considered \eg in \cite{Sorkin:2012sn,Coser:2017dtb,2004PhRvA..70e2329B,Arias:2018tmw}. %\comment{blah blah}

 Once again, we extend this bosonic algebra with $2N$ additional operators
 \begin{equation}
 \tilde f_i\equiv J f_i J\, .
 \end{equation}
These satisfy the commutation relations $[\tilde f_i, \tilde f_j]= -i C_{ij} $. Again, with this definition, the scalar correlators depend only on the density matrix of the original Hilbert space. In the purified state $\ket{\Omega}$ in the full space, we have
\begin{align}
\braket{\Omega | f_{i_1}\cdots f_{i_k} \tilde f_{j_1} \cdots \tilde f_{j_l} |\Omega}&=\braket{\Omega | f_{i_1}\cdots f_{i_k} J f_{j_1} \cdots  f_{j_l} |\Omega} \\ &=\braket{\Omega | f_{i_1}\cdots f_{i_k} \Delta^{1/2} f_{j_l} \cdots  f_{j_1} | \Omega} \\ &=\tr \left(\rho^{1/2} f_{i_1}\cdots f_{i_k}\rho^{1/2}  f_{j_l} \cdots  f_{j_1} \right)\, .
\end{align}
Let us denote 
\begin{equation}\label{Fij}
F_{ij}\equiv \braket{f_i f_j} \, .
\end{equation}
Organizing the operators in a single field $\mathcal{F}_i\equiv f_i$, $i=1,\dots,N$ and $\mathcal{F}_{i+N}\equiv \tilde f_i$, $i=1,\dots,N$, we can define the matrix of commutators
\begin{equation}\label{gammaa3}
\mathcal{C}_{ij}\equiv -i[\mathcal{F}_i,\mathcal{F}_j] \quad  \Rightarrow \quad
\mathcal{C}=  \left(
\begin{array}{cc}
 C & 0 \\
 0 &-C
\end{array}
\right)\, .
\end{equation}
Using the Hermiticity of the $\mathcal{F}_i$ it is easy to prove that\footnote{Note that when we write things like ${\rm Im} A_{ij}$, we literally refer to the matrix built from the imaginary parts of the components of the original matrix (and the same for the real parts).}
\begin{equation}\label{CF}
\mathcal{C}_{ij}=2\,  {\rm Im}\, \mathcal{F}_{ij}\, ,
\end{equation}
where we defined the matrix of correlators
\begin{equation}
\mathcal{F}_{ij}\equiv \braket{\Omega| \mathcal{F}_i \mathcal{F}_j  |\Omega}\, , \quad i=1,\dots,2N\, .
 \end{equation}
The different blocks in this matrix turn out to be given by
  \begin{align} \label{phiii2}
\mathcal{F}&=\left(
\begin{array}{cc}
 F &   i CV g(V^2 )  \\
 i CV g(V^2 ) & F-iC
\end{array}
\right)\, , \quad \text{where} \quad V\equiv -i C^{-1} F -\frac{1}{2}\, , %= -\frac{i}{2} {\rm I}^{-1}{\rm R}\, ,
\end{align}
 and $g(A)$ was defined in \req{g}. This matrix can also be written as
 %. This can be written fully in terms of $F$ as
  \begin{align} \label{F113}
\mathcal{F}&=\left(
\begin{array}{cc}
 {\rm R}+ i {\rm I}  &  g\left(-\frac{1}{4} {\rm R} {\rm I}^{-1}{\rm R}  {\rm I}^{-1}  \right) {\rm R} \\
 g\left(-\frac{1}{4} {\rm R} {\rm I}^{-1}{\rm R}  {\rm I}^{-1}  \right) {\rm R} &  {\rm R}- i {\rm I}
\end{array}
\right)\, ,
\end{align}
where we defined $ {\rm R}_{ij}\equiv  {\rm Re}(F_{ij})$, $ {\rm I}_{ij}\equiv {\rm Im}(F_{ij})$.
 Note that the off-diagonal terms are manifestly real. In order to evaluate the entropy associated to some region $Y$, we define the matrix
\begin{equation}\label{vy}
\mathcal{V}_Y \equiv  -i  ( \mathcal{C}_Y)^{-1} \mathcal{F}_Y -\frac{1}{2}\, ,
\end{equation}
where $\mathcal{C}_Y$ and $\mathcal{F}_Y$ are the restrictions of $\mathcal{C}$ and $\mathcal{F}$ to $Y$. Then, the 
corresponding Von Neumann entropy can be obtained as
\begin{equation}\label{svn}
S(Y)=%\tr (V+1/2) \log |V+1/2| =
\tr \left[ (\mathcal{V}_Y+1/2)\log |\mathcal{V}_Y+1/2| \right]\, .
\end{equation}
%where $\Theta(x)$ is the Heaviside theta function.
%The explicit form of this matrix can be obtained using \req{phiii2} and reads
%\begin{align}\label{vcal}
%\mathcal{V}=  
%\left(\begin{array}{cc}   V    &   -\sqrt{V^2-1/4}  \\   \sqrt{V^2-1/4}  &  -V   \end{array}\right)\, , \quad \text{where} \quad V= -\frac{i}{2} {\rm I}^{-1}{\rm R} \, .
%\end{align}
When computing reflected entropies, we need to evaluate $\mathcal{C}$ and $\mathcal{F}$ for all sites belonging to $A$ and $B$, and then obtain the restrictions of their different blocks to the region $A$.  

As a check of our results, we can observe that one should find $S(Y)=0$ when applied to the global state, which means that the unrestricted matrix $\mathcal{V}$ should be such that $\mathcal{V}^2=1/4$, which can be easily verified to be the case. Observe also that, once again, these expressions can be seen as generalizations of the analogous entanglement entropy formulas. For that quantity \req{svn} holds \cite{Sorkin:2012sn} with $ \mathcal{V}_Y$ replaced by $V_Y\equiv -i  ( C_Y)^{-1} F_Y -\frac{1}{2}$.

 The terms appearing in the diagonal of $\mathcal{F}$ in the expressions above follow straightforwardly, but the origin of the off-diagonal pieces requires some further explanation. In order to see where they come from, let us define vectors $\vec f\equiv (f_1, \dots,f_{2N})^T$ and $\vec \Phi\equiv (\phi_1,\dots,\phi_N,\pi_1,\dots,\pi_N)^T $ and $\vec {\tilde f}\equiv (\tilde f_1, \dots, \tilde f_{2N})^T$ and $\vec {\tilde \Phi}\equiv (\tilde \phi_1,\dots,\tilde \phi_N, \tilde \pi_1,\dots,\tilde \pi_N)^T $. As argued in \cite{Arias:2018tmw}, we can perform a change of basis to relate the $\vec f$ and $\vec \Phi$ representations as
%On the other hand, we can use the change of basis 
$\vec \Phi=Q O \vec f$ where $Q={\rm diag}(D^{-1/2},D^{-1/2})$ being $D$ a diagonal matrix with positive elements, and $O$  an orthogonal matrix.
  On the one hand, we have
 \begin{align}
&C=O^T Q^{-1}  \left(\begin{array}{cc} 0 & 1\\ -1 & 0 \end{array}\right) Q^{-1} O\, ,  \quad F=O^T Q^{-1}\left(\begin{array}{cc}  X & i/2 \\ -i/2 & P \end{array}\right) Q^{-1} O\, , \\  & \Rightarrow  V=O^TQ \left(\begin{array}{cc}  0 & i P \\ -iX & 0 \end{array}\right)Q^{-1}O\, , \quad F-iC= O^T Q^{-1}\left(\begin{array}{cc}  X & -i/2 \\ i/2 & P \end{array}\right) Q^{-1} O\, .
 \end{align}
Now, our goal is to evaluate $\braket{f_i \tilde f_j}$. In order to do that, we use the result obtained in \req{phiii} in the $\phi,\pi$ basis. We have
\begin{equation}\label{phiphi} 
\braket{\Phi \tilde \Phi}= \left(\begin{array}{cc} \braket{\phi \tilde \phi} & 0\\ 0 & \braket{\pi \tilde \pi} \end{array}\right)=   \left(\begin{array}{cc} g(XP) X & 0\\ 0 & g(PX) P \end{array}\right)\, .
\end{equation}
Then, we have
\begin{equation}
\braket{\Phi \tilde \Phi}= Q O \braket{f \tilde f}  O^T Q \quad \Rightarrow \quad \braket{f \tilde f} = O^T Q^{-1} \braket{\Phi \tilde \Phi} Q^{-1} O\, .
\end{equation}
Now, in order to write the expression in \req{phiphi} in terms of correlators of $f_i$, we can use the above expressions for $C$ and $V$. We find %matrix
%
 %$V$ defined in \req{phiii2} and
%\begin{equation}
%V^2= O^T Q  \left(\begin{array}{cc} PX & 0\\ 0 & XP  \end{array}\right) Q^{-1} O\, .
%\end{equation}
%We find
\begin{equation}
i  CV g( V^2 ) =O^T Q^{-1} \left(\begin{array}{cc} g(XP) X & 0\\ 0 & g(PX) P \end{array}\right) Q^{-1} O \quad \Rightarrow \quad \braket{f \tilde f}=i  CV g( V^2 ) \, ,
\end{equation}
which is the desired relation appearing in the off-diagonal blocks of  $\mathcal{F}$.

\section{Reflected entropy for a $d=2$ chiral scalar}\label{chiralsca}
In this section we evaluate numerically the reflected entropy for two intervals  for a chiral scalar field as a function of the conformal   cross-ratio and compare the result to the ones corresponding to holographic Einstein gravity and a free fermion. We also study the eigenvalues spectrum of the matrix of correlators which intervenes in the computation of the reflected entropy and comment on its differences with respect to the one required for the evaluation of the usual type-III entanglement entropy of a single interval. We also verify the monotonicity of reflected entropy under inclusions both for the scalar and the fermion.

 \subsection{Reflected entropy for two intervals}
 %Our goal in this section is to compute the reflected entropy for two disjoint interval regions $A$, $B$  
 The lattice Hamiltonian for a chiral scalar in $1+1$ dimensions can be taken to be
 %For a lattice Hamiltonian of the form
 \begin{equation}
 H=\frac{1}{2}\sum_i f_i^2\, .
 \end{equation}
In this case, the correlator defined in \req{Fij} was obtained in \cite{Arias:2018tmw}, the result being
 \begin{equation}\label{fff}
 F_{ij}= \begin{cases} -\frac{1+(-1)^{i-j}}{\pi ((i-j)^2-1)}  \, , \quad & |i-j|\neq 1 \, , \\ +\frac{i}{2} \left( \delta_{j i+1}-\delta_{j i-1}\right) \, , \quad & |i-j|= 1 \, .
 \end{cases}
 \end{equation}

 Given two regions $A$ and $B$, we can evaluate the reflected entropy as the von Neumann entropy of $\rho_{AA^*}$ using the expression for $F_{ij}$ above and the formulas obtained in the previous section. The indices $i,j$ in $F_{ij}$  take values in sites belonging to the region $A\cup B$. Namely,   if we define the discretized intervals through $A\cup B = (a_1, a_{1}+1, \dots  , b_{1}-1, b_1)\cup (a_2, a_2+1, \dots , b_2-1, b_2)$, then  $i,j$ take  values $j = a_1, a_1+1, \dots  , b_1-1, b_1, a_2, a_2+1, \dots ,  b_2-1, b_2$. Given $(a_1, b_1)$ and $(a_2, b_2)$ as input, which determine the length and separation of the corresponding intervals, we can then evaluate the matrix $F_{ij}$. The real and imaginary parts of its components are easily obtained from \req{fff} and given by
 \begin{equation}
 {\rm Re}\, F_{ij}= \begin{cases} -\frac{1+(-1)^{i-j}}{\pi ((i-j)^2-1)}  \, , \quad & |i-j|\neq 1 \, , \\ 0 \, , \quad & |i-j|= 1 \, ,
 \end{cases} \quad  
 {\rm Im}\, F_{ij}= \frac{1}{2} \left( \delta_{j i+1}-\delta_{j i-1}\right)  \, .
 % ({\rm Im} F)_{ij}= \begin{cases} 0  \, , \quad & |i-j|\neq 1 \, , \\ +\frac{1}{2} \left( \delta_{j i+1}-\delta_{j i-1}\right)  \, , \quad & |i-j|= 1 \, .
% \end{cases}
 \end{equation}
 With these matrices at hand, we can numerically compute the diagonal terms appearing in $\mathcal{C}$ and $\mathcal{F}$ in \req{gammaa3} and \req{F113} respectively, as well as the combination $W\equiv (-\tfrac{i}{2}{\rm R} {\rm I }^{-1})_{ij}$, required for the off-diagonal blocks of $\mathcal{F}$. In order to obtain those, we first diagonalize $W$. Given its eigenvalues $\{ d_m\}$, we build the diagonal matrix $|d_m |^{-1} \sqrt{d_m^2-1/4}\,  \delta_{mn}$ and transform it back to the original basis, which yields $\left.g(-\tfrac{1}{4} {\rm R} {\rm I}^{-1} {\rm R} {\rm I}^{-1})\right|_{ij}$. Multiplying by ${\rm R}$, we obtain the off-diagonal blocks of $\mathcal{F}$. Using these matrices we can obtain the von Neumann entropy associated to $\rho_{AA^*}$, from
the submatrices corresponding to the $A$ sites. These correspond to the first $(b_1 - a_1) \times (b_1 -a_1)$-dimensional blocks in each case. With those pieces we can finally build the matrices $\mathcal{C}|_{AA^*}$ and $\mathcal{F}|_{AA^*}$ as
%\left.\mathcal{V}\right|_A=  
%\frac{i}{2}\left(\begin{array}{cc}  \left.- {\rm I}^{-1}{\rm R} \right|_A  &  \left.-  g (-\frac{1}{4} {\rm I}^{-1}{\rm R}{\rm I}^{-1}{\rm R}) {\rm I}^{-1}{\rm R} \right|_A \\ \left.  g (-\tfrac{1}{4} {\rm I}^{-1} {\rm R}{\rm I}^{-1}{\rm R}) {\rm I}^{-1}{\rm R} \right|_A & \left. {\rm I}^{-1}{\rm R} \right|_A  \end{array}\right)\, .
%\end{align}
%\begin{align}\label{vcal2}
%\left.\mathcal{V}\right|_A=  
%\left(\begin{array}{cc}  \left. M \right|_A   &  \left. \sqrt{M^2-1/4} \right|_A \\ \left.  -\sqrt{M^2-1/4} \right|_A & \left. -M \right|_A  \end{array}\right)\, , \quad \text{where} \quad M\equiv -\frac{i}{2} {\rm I}^{-1}{\rm R} \, .
%\end{align}
  \begin{align} \label{F1134}
\left. \mathcal{F} \right|_{AA^*}&=\left(
\begin{array}{cc}
\left. [{\rm R}+ i {\rm I}] \right|_A  &  \left.\left[ g\left(-\frac{1}{4} {\rm R} {\rm I}^{-1}{\rm R}  {\rm I}^{-1}  \right) {\rm R} \right]\right|_A\\
 \left.\left[ g\left(-\frac{1}{4} {\rm R} {\rm I}^{-1}{\rm R}  {\rm I}^{-1}  \right) {\rm R} \right]\right|_A & \left. [{\rm R}- i {\rm I}]\right|_A 
\end{array}
\right)\, , \quad \left. \mathcal{C} \right|_{AA^*}&=\left(
\begin{array}{cc}
\left. 2 {\rm I} \right|_A  & \left.0\right|_A\\
 \left.0\right|_A & -\left. 2 {\rm I}\right|_A 
\end{array}
\right)\, .
\end{align}
The last step is to evaluate $\mathcal{V}_{AA^*} \equiv  -i  ( \mathcal{C}|_{AA^*})^{-1} \mathcal{F}|_{AA^*} -\frac{1}{2}$.
Denoting its eigenvalues as $\{\nu_m\}$, the reflected entropy can be finally obtained from \req{svn} as 
 \begin{equation}\label{eigenR}
 R_{\rm scal.}=\sum_{m} (\nu_m+1/2)\log | \nu_m+1/2|\, .
\end{equation} 
Lattice calculations give rise to a doubling of eigenvalues, so when showing results we need to divide the numerical results by 2. On the other hand, from now on we will normalize reflected entropies by the central charge $c$ of the corresponding theory, which in the case of the chiral scalar is $c=1/2$. Hence, the numerical results obtained following the above procedure automatically yield $ R_{\rm scal.}/c$.

\begin{figure}[t] \centering
	\includegraphics[scale=0.75]{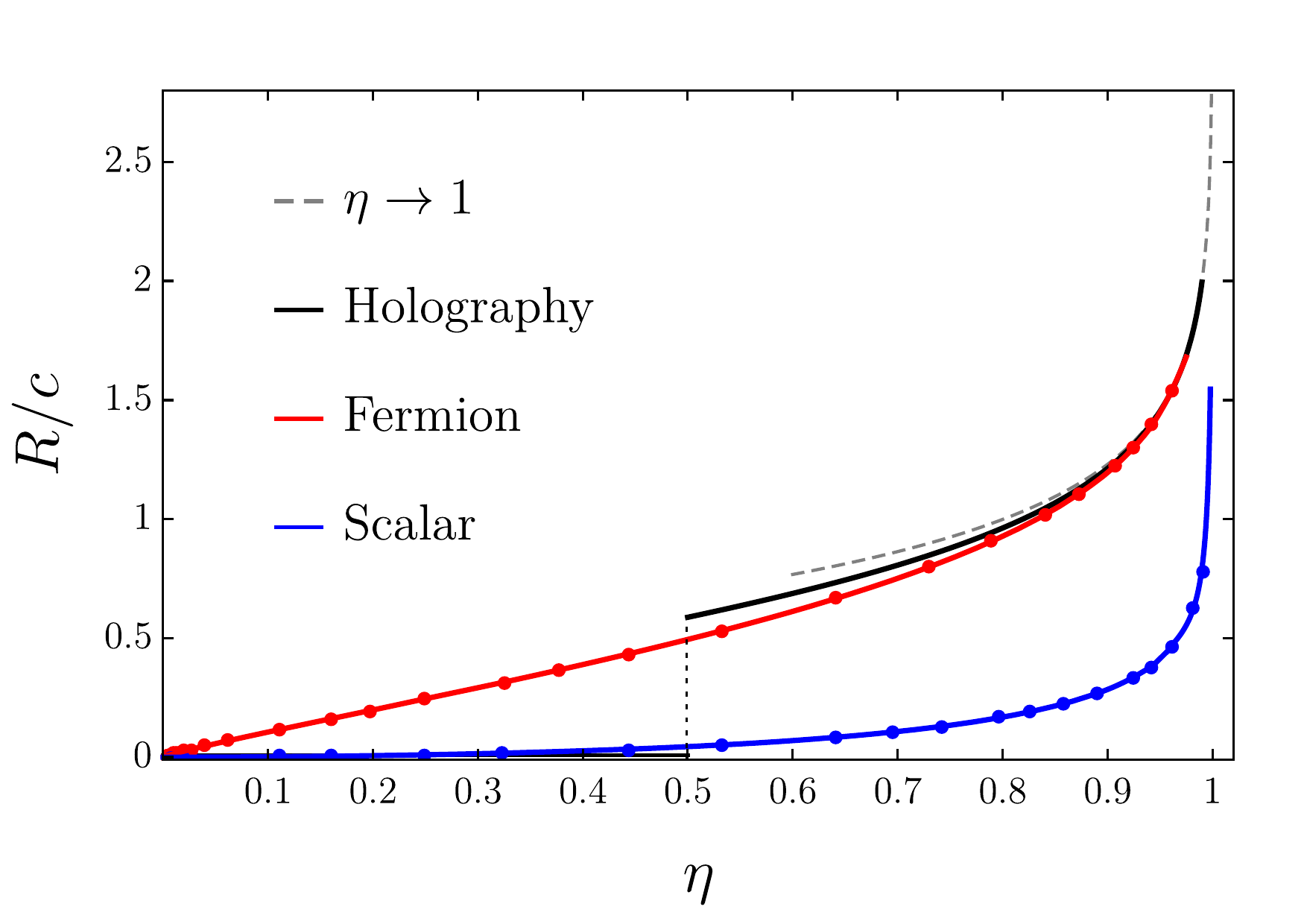}
	\caption{We plot the reflected entropy normalized by the central charge, $R/c$,   as a function of the cross-ratio $\eta$ for: a chiral scalar (blue line and dots), a free fermion (red line and dots) \cite{Bueno:2020vnx} and holographic Einstein gravity (black line) \cite{Dutta:2019gen}. The latter corresponds to the leading-order result in the Newton constant which drops to zero for $\eta =1/2$. The gray dashed line is the general-theory  behavior  as $\eta \rightarrow 1$.    }
	\label{refiss1}
\end{figure}

In the continuum, the reflected entropy for two intervals of lengths $L_A$, $L_B$ separated a distance $\ell$ is a function of the conformal cross-ratio
\begin{equation}\label{cr7}
\eta \equiv \frac{(b_1-a_1)(b_2-a_2)}{(a_2-a_1)(b_2-b_1)}=\frac{L_A  L_B}{(\ell+L_A)(\ell + L_B)} \, .
\end{equation}
In order to obtain $ R_{\rm scal.}(\eta)$ in that limit, we fix $\eta$ and consider an increasing number of points in the discretized intervals. The results for the reflected entropy asymptote to their continuum values, which we obtain through a polynomial fit in the inverse size of the  intervals. We plot our results in Fig. \ref{refiss1}. In the same plot, we include the results corresponding to holographic Einstein gravity and a free fermion. The former was obtained in \cite{Dutta:2019gen} using replica-trick techniques, and reads
\begin{equation}
R_{\rm holo.}(\eta)= \begin{cases} 
\frac{2c}{ 3} \log \left[\frac{1+\sqrt{\eta}}{\sqrt{1-\eta}} \right]    +  \mathcal{O}(c^0)\, , \quad \text{ for } \quad \eta>1/2\, , \\
\mathcal{O}(c^{0}) \, , \quad   \quad \quad \quad \quad \,    \,   \,    \,   \, \,  \, \, \,   \, \,  \,   \quad  \text{ for } \quad \eta<1/2\, .
\end{cases}
\end{equation}
This agrees with previous $E_W$ calculations  \cite{Takayanagi:2017knl,Nguyen:2017yqw}. On the other hand, the fermion results were obtained using numerical methods in \cite{Bueno:2020vnx}. In Fig. \ref{refiss1} we have also included the  $\eta\rightarrow 1$ limit which was argued to hold for general $d=2$ CFTs in \cite{Dutta:2019gen}. This reads
\begin{equation}\label{r11}
R(\eta\rightarrow 1)=-\frac{c}{3}\log (1-\eta)+\frac{c}{3}\log 4\, .
\end{equation}
While the fermion and holographic results clearly approach the limiting curve in the expected regime (doing so from below), the scalar takes values which are considerably smaller for values of $\eta$ very close to one. In appendix \ref{ape} we present the numerical values of the data points shown in Fig.\,\ref{refiss1} both for the scalar and the fermion, which may be useful for future comparisons.

\begin{figure}[t]\hspace{-0.3cm}
    \includegraphics[scale=0.67]{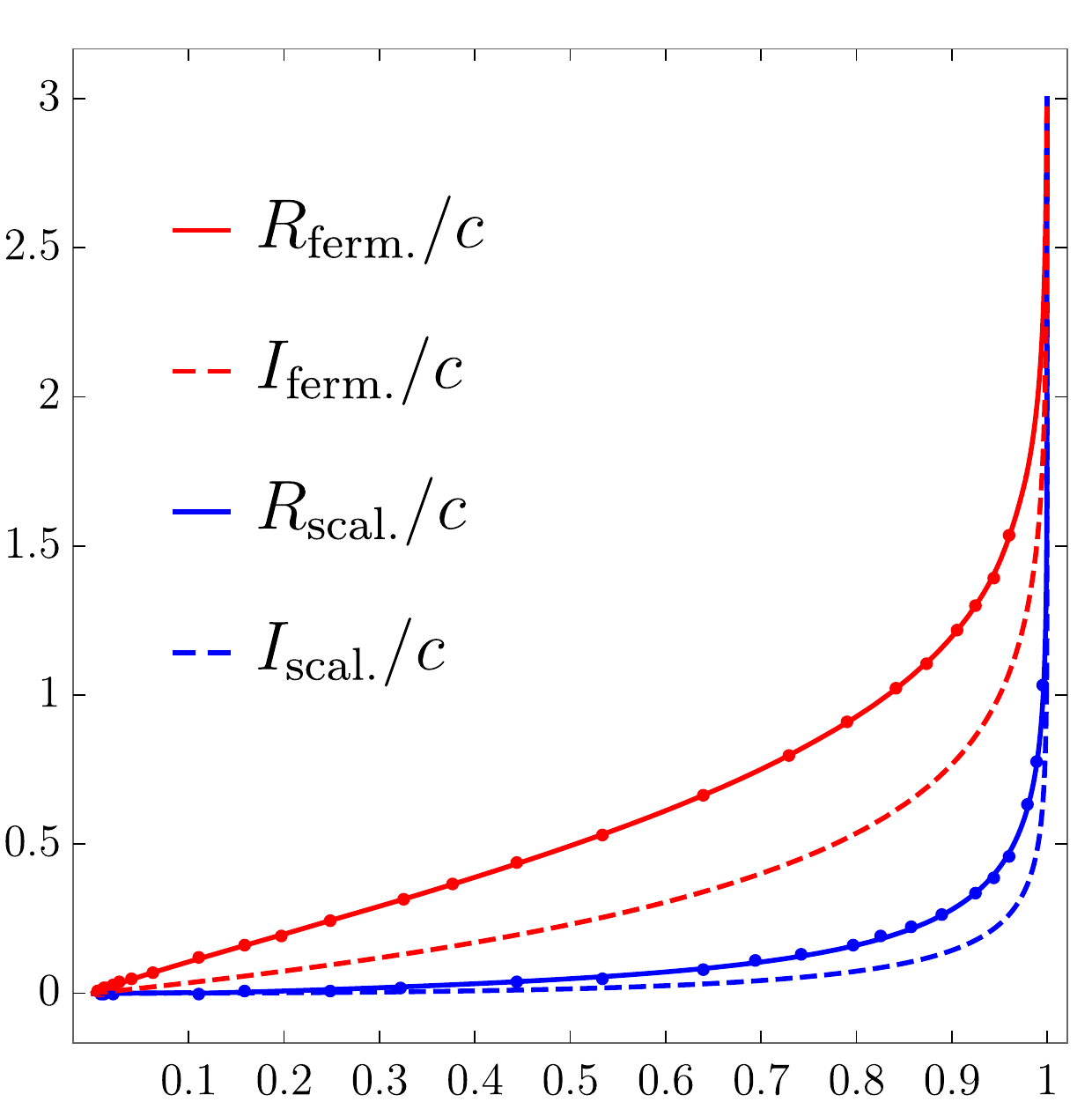}
	\includegraphics[scale=0.645]{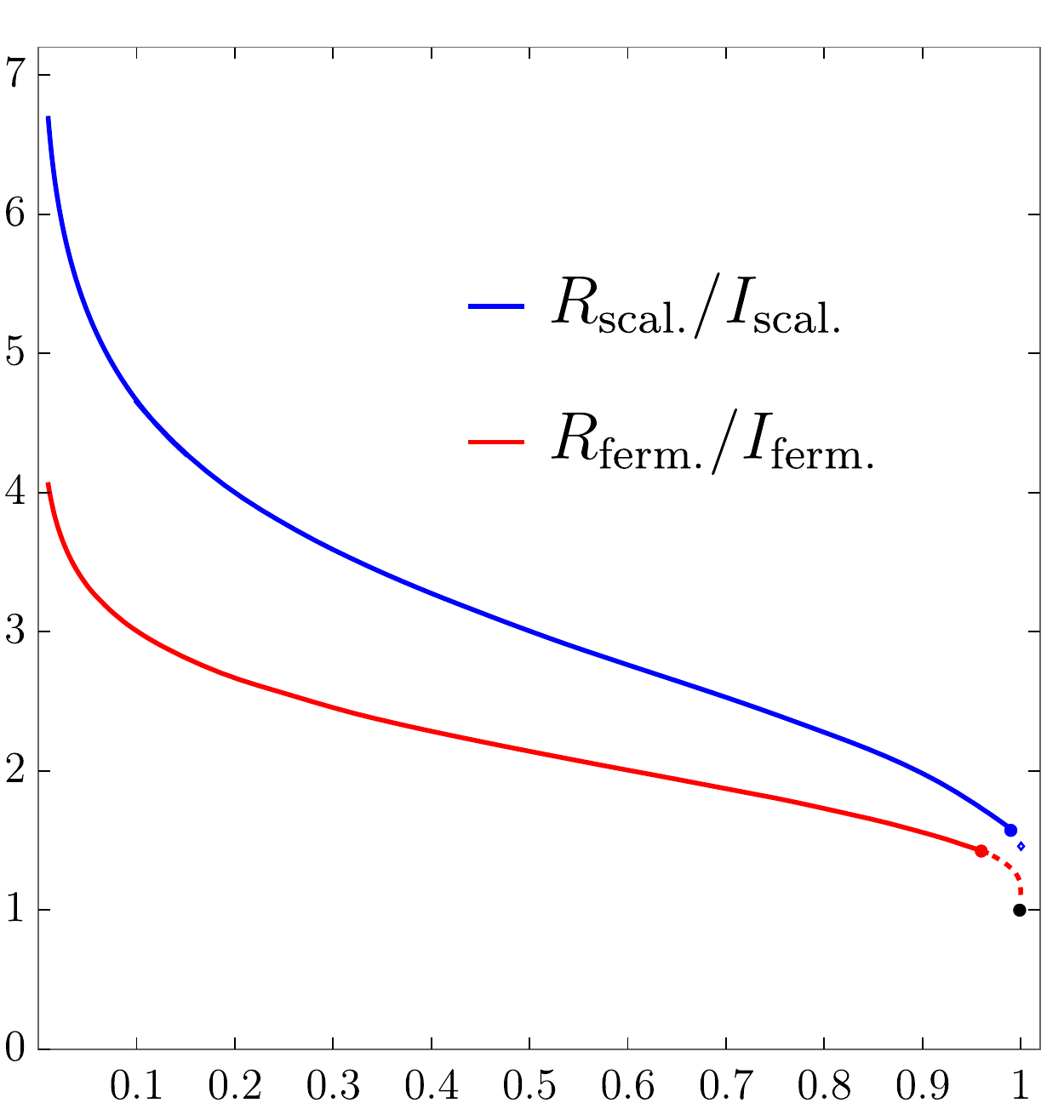}
	\caption{We plot the reflected entropy and mutual information for two intervals $A,B$, as a function of the cross-ratio for a free fermion (red) and a free scalar (blue). In the right we plot the quotient between both quantities for each model. The black dot corresponds to the limit $\eta=1$, where both quotients should tend to one. The red and blue dots correspond to the greatest values of $\eta$ for which we numerically evaluated the reflected entropy for each model. The red dotted line has been computed using the general-CFT formula \req{r11} and \req{mutuss} and is valid for $\eta\rightarrow 1$. In the case of the scalar, the curve becomes very steep near $\eta\rightarrow 1$ because of $U(\tau)$. For instance, the small blue diamond shown in the figure corresponds to the value $\eta=0.9999999999999999$, for which $R_{\rm scal.}/I_{\rm scal.}=1.470488$.    }
	\label{refiss12222}
\end{figure}

In spite of being much smaller than the fermion and holographic results, we can verify that $R_{\rm scal.}$ is indeed greater than the mutual information $I_{\rm scal.}$ as required by the general inequality in \req{ir}. For that, we recall  the  results for the mutual information of fermion and scalar \cite{Arias:2018tmw}. These are given by
\begin{equation}\label{mutuss}
I_{\rm ferm.}/c=-\frac{1}{3}\log (1-\eta)\, , \quad I_{\rm scal.}/c=-\frac{1}{3}\log (1-\eta)+2U(\eta)\, ,
\end{equation} 
where 
\begin{equation}
U(\eta)\equiv  -\frac{i\pi}{2}   \int_0^{\infty} ds \frac{s}{\sinh^2(\pi s)} \log \left[\frac{_2F_1[1+is,-is; 1; \eta]}{_2F_1[1-is,+is; 1; \eta]} \right] \, ,
\end{equation}
which is a real and negative function for all values of $\eta$. We plot the corresponding reflected entropies and mutual informations for both models in Fig.\,\ref{refiss12222}. In both cases, the inequality is satisfied, as it should, and the quotient $R/I$ monotonously decreases for growing values of $\eta$. In the limit $\eta\rightarrow 1$, both quotients tend to one. In the case of the scalar, it requires values of $\eta$ extremely close to one to approach that limit ---see blue diamond in the right plot. This is related to the behavior of the function $U(\eta)$, which goes as $U(\eta) \sim -\tfrac{1}{2} \log \left[ -\log [1-\eta]\right]$ for $\eta\rightarrow 1$ \cite{Arias:2018tmw}.
In the opposite limit, \ie for $\eta\rightarrow0$, the quotients seem to diverge logarithmically. In the case of the fermion, we found that the tentative function \cite{Bueno:2020vnx}
\begin{equation}
R_{\rm ferm.}(\eta\rightarrow 0)/c \sim -0.15\eta \log \eta+ 0.67 \eta +\dots
\end{equation}
fits reasonably well the numerical data for values of the cross ratio $\eta\lesssim 0.1$. In the case of the scalar, a similar analysis  suggests that the leading order term takes the form
\begin{equation}\label{rc0}
R_{\rm scal.}(\eta\rightarrow 0)/c \sim -0.04\eta^2 \log \eta+\dots
\end{equation}
The fit in this case goes wrong much faster than in the case of the fermion, and can only be trusted for values of the cross ratio  $\eta\lesssim 0.001$. In spite of this limited range of validity, we are rather confident the functional dependence of the leading term is the one shown in \req{rc0}. In the case of the mutual informations, one finds instead \cite{Casini:2009vk,Casini:2005rm,Arias:2018tmw}
\begin{align}
I_{\rm ferm.}(\eta\rightarrow 0)/c   \sim  \frac{1}{3}\eta+ \dots   \, , \quad
I_{\rm scal.}(\eta\rightarrow 0)/c  \sim  \frac{1}{30}\eta^2+ \dots \, .
\end{align}
These results reflect the different nature of both quantities. While mutual information admits a power-law expansion in that limit \cite{Calabrese:2009ez,Calabrese:2010he,Cardy:2013nua,Agon:2015ftl}, which reflects the fact that it measures correlations between operators exclusively localized in $A,B$, the information captured by reflected entropy is in fact spread throughout the whole real line (except for the region corresponding to the interval $B$). The latter fact was shown very explicitly in the case of the fermion in \cite{Bueno:2020vnx}, where a notion of spatial-density for the corresponding type-I algebra was introduced.   %which can be understood as a consequence of the existence of an operator product expansion of twist operators localized on the entangling surface in the replica trick \cite{} \comment{mooore }

 \subsection{Eigenvalues spectrum}
In \cite{Bueno:2020vnx}, we studied how the spectra of the correlator matrices entering the entanglement  and reflected entropies differed from each other for a $(1+1)$-dimensional free fermion. The goal of this subsection is to perform an analogous analysis in the case of the chiral scalar. Just like for the fermion, the formulas required for the evaluation of reflected entropy in the case of free scalars are also identical to the entanglement entropy ones ---namely, they have the same form in terms of certain two-point functions of the fields. The difference between both quantities is that in the entanglement entropy case the relevant matrices are $C_A$ and $F_A$, whereas for the reflected we need $\mathcal{C}_{AA^*}$ and $\mathcal{F}_{AA^*}$. In this setup, this is what makes the difference between computing a von Neumann entropy for a type-III algebra associated to region $A$, and a von Neumann entropy for the canonical type-I algebra associated to regions $A$ and $B$, \ie a reflected entropy. 

\begin{figure}[t]\hspace{-0.3cm}
    \includegraphics[scale=0.68]{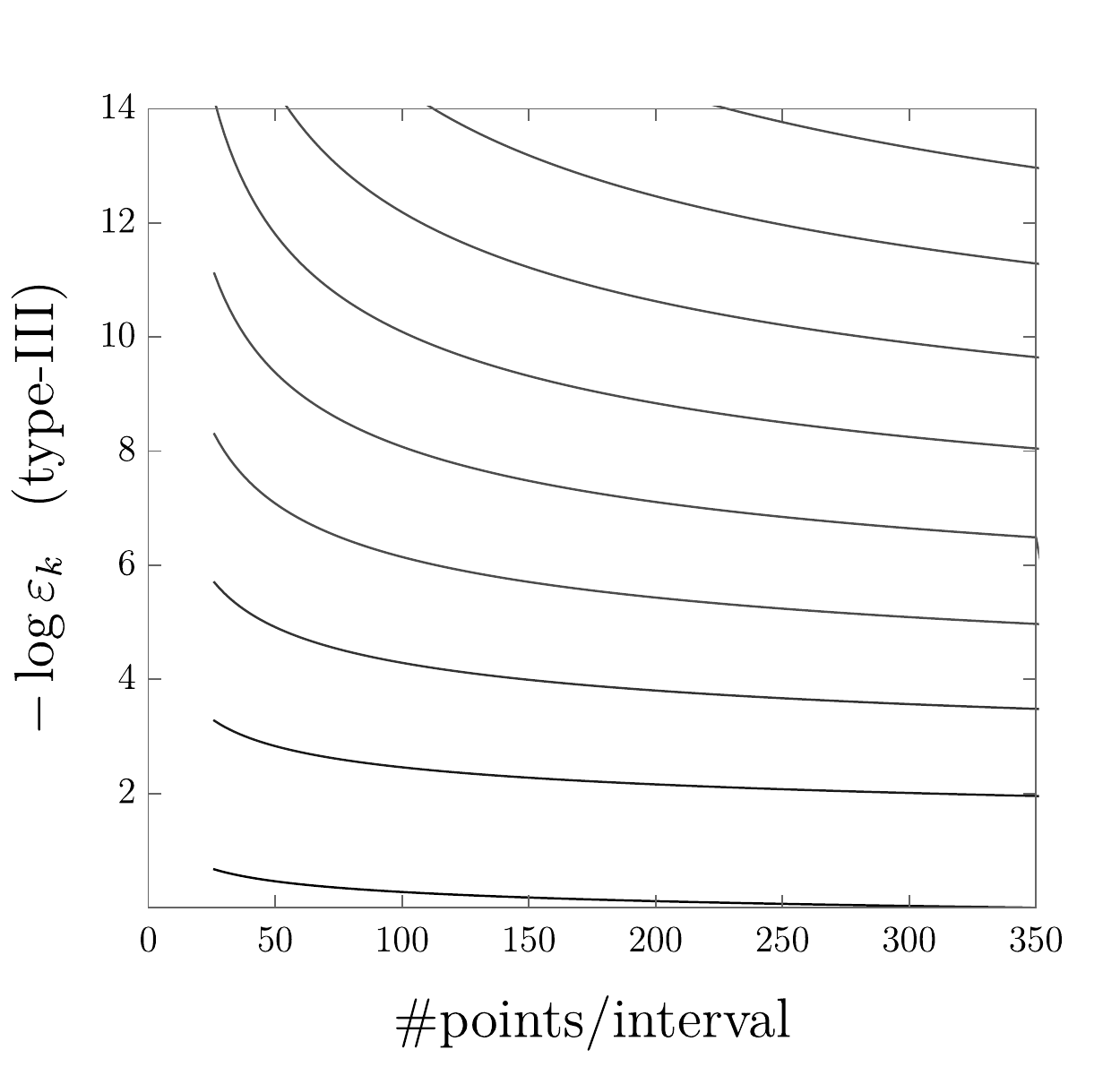}
	\includegraphics[scale=0.68]{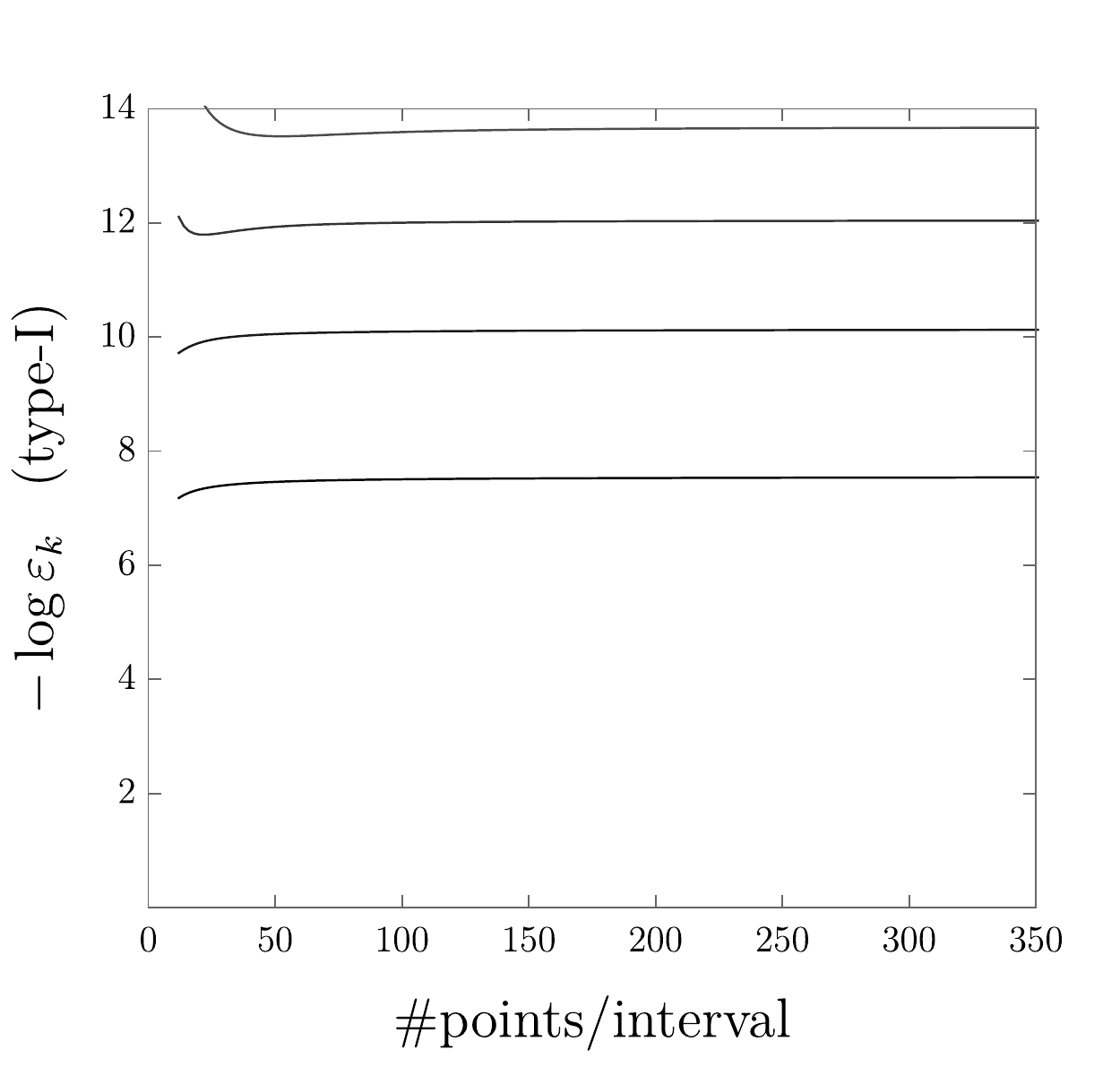}
	\caption{We plot the ``leading'' eigenvalues of the correlator matrices  $V_A$ and $\mathcal{V}_{AA^*}$  involved in the evaluation of: the usual type-III entanglement entropy for a single interval (left); the reflected entropy $R(A, B)$ for two invervals $A$,$B$ with cross-ratio $\eta=1/4$ (right). For both plots, the horizontal axis corresponds to the number of points taken for the intervals ($A$ in the first case and both $A$ and $B$ in the second). In both cases, we use a logarithmic function of the eigenvalues which simplifies presentation of several eigenvalues in the same plot ---see \req{arraa}.  }
	\label{refiss122}
\end{figure}

The eigenvalues of $\mathcal{V}_{AA^*}$ always appear doubled, as mentioned above. In the following discussion we just remove the repeated eigenvalues and multiply the result by 2. For each remaining eigenvalue $\nu_j$ there is always another one corresponding to $-\nu_j$. Hence, it is useful to arrange the eigenvalues as 
\begin{equation}\label{arraa}
\nu_{2k}\equiv \tfrac{1}{2}+\varepsilon_k\, , \quad \nu_{2k-1}\equiv-\tfrac{1}{2}-\varepsilon_k\, , \quad \text{with} \quad k=1,2,\dots,\#_A\, ,
\end{equation}
where the $\varepsilon_k$ are positive numbers and $\#_A$ is the number of lattice points corresponding to the interval $A$. The continuum limit corresponds to $\#_A\rightarrow \infty$. The above expressions can be inverted as $\varepsilon_k=(\nu_{2k}- \nu_{2k-1}-1)/2=\nu_{2k}-1/2=-\nu_{2k-1}-1/2$.
Then, we can rewrite the reflected entropy  \req{eigenR} as
 \begin{equation}
 R_{\rm scal.}=2\sum_{k=1}^{\#_A}\left[  (\varepsilon_k+1) \log (1+\varepsilon_k) - \varepsilon_k \log \varepsilon_k  \right] \, .
\end{equation} 
Except for values of $\eta$ very close to $1$, the $\varepsilon_k$ are all very small numbers, so $R_{\rm scal.}$ is approximately given by
 \begin{equation}
 R_{\rm scal.}=2\sum_{k=1}^{\#_A} \left[ \varepsilon_k [1-\log \varepsilon_k ] +  \frac{\varepsilon_k^2}{2} + \mathcal{O}( \varepsilon_k^3) \right] \, .
\end{equation} 
In this expression, both $ \varepsilon_k$ and $-\varepsilon_k \log \varepsilon_k$ make comparable contributions to $ R_{\rm scal.}$ for the most relevant eigenvalues, but $-\varepsilon_k \log \varepsilon_k$ always dominates whenever $\varepsilon_k <1/ e$, which again is the case for all values of $\eta$ except for those extremely close to $\eta=1$. In order to compare the behavior of the eigenvalues of $\mathcal{V}_{AA^*}$ with those of $V_A$ we choose to plot $-\log\varepsilon_k$ as a function of the number of points in the interval $A$. Note that the smaller the values of $-\log\varepsilon_k$ for a given pair of eigenvalues $\{\nu_{2k},\nu_{2k-1}\}$, the greater the contribution to $R_{\rm scal.}$, since the resulting function appears multiplied by $\varepsilon_k$ in the reflected entropy expression. Indeed, the closer to $0$ a given $\varepsilon_j$ is, the smallest its contribution, since then $\varepsilon_j \log \varepsilon_j \rightarrow 0$, and $(\varepsilon_j+1) \log (\varepsilon_j+1) \rightarrow \log 1=0$.   As for the eigenvalues of $V_A$, in that case there is no doubling but, just like for the reflected entropy, for each positive eigenvalue there always appears its negative version, so the arrangement \req{arraa} can be performed as well, where now the $\varepsilon_k$ are no longer small in general.

In Fig.\,\ref{refiss122} we plot the function $-\log\varepsilon_k$ as we approach the continuum for the eigenvalues of $\mathcal{V}_{AA^*}$ and $V_A$ which contribute the most to the reflected and entanglement entropies, respectively. The greatest contribution comes, in both cases, from the lowest curve, and so on. In a very similar fashion to the situation encountered for a free fermion in \cite{Bueno:2020vnx}, we observe that only a few eigenvalues make a significant contribution to $R(A,B)$. The eigenvalues  quickly stabilize as we approach the continuum, as expected for a finite type-I algebra. On the other hand, in the entanglement entropy case, an increasing number of eigenvalues of $V_A$ become relevant, which produces the usual logarithmically  divergent behavior. 

It is natural to wonder how well the first eigenvalues manage to reproduce the full reflected entropy result. In order to test this, one can define ``partial'' reflected entropies as
 \begin{equation}
 R_{\rm scal.}^{(p)}=2\sum_{k=1}^{p}\left[  (\varepsilon_k+1) \log (1+\varepsilon_k) - \varepsilon_k \log \varepsilon_k  \right] \, ,
\end{equation} 
where again it is understood that we have arranged the $\varepsilon_k$ from greatest to smallest. For our working example of $\eta=1/4$, one finds,
\begin{align}
 R_{\rm scal.}^{(1)}(1/4)&=0.0089725\, , \\
 R_{\rm scal.}^{(2)}(1/4)&= 0.0098531  \, , \\
  R_{\rm scal.}^{(3)}(1/4)&= 0.0100063  \, , \\
   R_{\rm scal.}^{(4)}(1/4)&= 0.0100385  \, , \\
  R_{\rm scal.}^{(\infty)}(1/4)&=0.0100512 \, .
\end{align}
As we can see, already with four eigenvalues we obtain a pretty accurate approximation to the full answer. A similar situation is encountered for intermediate values of $\eta$. On the other hand, as we approach the $\eta \rightarrow 1$ limit, a growing number of eigenvalues is required.

%\comment{comment on what eigenvalues contribute the most, those close to $\nu_m=0$, small contribution from those close to $\nu_m=-1/2,1/2$. Finite result for entropy associated to finite number of relevant eigenvalues , infinite entropy should be associated to growing number of eigenvalues with $|\nu_m|\neq 1/2$ as the continuum is approached. This is exactly the behavior we observed in the case of the $1+1$-dimensional free fermion. Notion of partial reflected entropies. }

 \subsection{Monotonicity of reflected entropy}
 
\begin{figure}[t]\centering
    \includegraphics[scale=0.68]{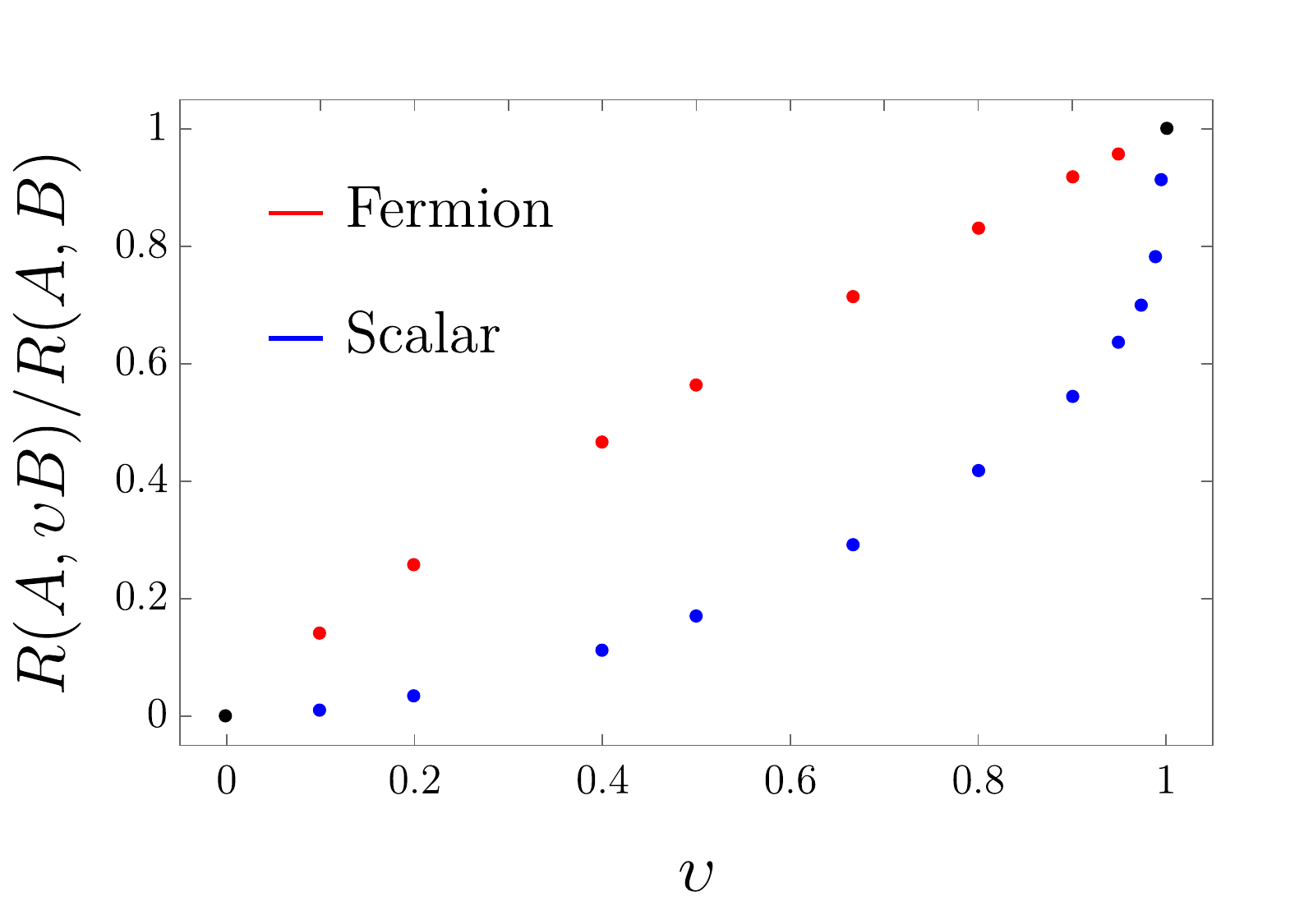}
	\caption{For a free fermion and a chiral scalar, we plot the reflected entropy corresponding to a fixed interval $A$ and a region $\upsilon B$ consisting of two intervals obtained as follows: given a single interval $B$ identical to $A$ and with a fixed cross-ratio $\eta=1/9$, we remove a certain subset of $B$ symmetric around its center so that we keep a total fraction $\upsilon$ of $B$. For instance, $\upsilon=2/5$ means that we have divided $B$ in five identical pieces and we have computed reflected entropy for $A$ and the pair of intervals resulting from removing the three intermediate fifths of $B$. The result appears normalized by $R(A,B)$, \ie by the one obtained by considering the full interval $B$.  The black dots correspond to the limit cases and are shared by the two models.     }
	\label{mono}
\end{figure}
The monotonicity of reflected entropy under inclusions (or its lack thereof) is an open problem. Namely, it is not know whether 
\begin{equation}\label{monos}
R(A,BC) \overset{?}{ \geq} R(A,B)\, ,
\end{equation} 
is a general property of reflected entropy. An analogous inequality was proven for integer-$n>1$ R\'enyi versions of the reflected entropy in  \cite{Dutta:2019gen}, but the $n=1$ case still remains uncertain.

We have tested the validity of \req{monos} for the free scalar and the free fermion by computing reflected entropy of pairs of regions $A$ and $\upsilon B$ where only a fraction $\upsilon$ of the original interval $B$ is considered. In Fig.\,\ref{mono}, we have considered a particular case corresponding to intervals $A,B$ with cross-ratio $\eta=1/9$.
We find that \req{monos} always holds, \ie as we increase the fraction of $B$ which we consider, the reflected entropy grows. Hence, reflected entropy indeed satisfies the monotonicity property in these cases. We have repeated the experiment for other values of $\eta$, and \req{monos} is always respected both for the scalar and the fermion. While our analysis is only partial, our results suggest that \req{monos} indeed holds for all possible choices of $A,B,C$ in the case of free scalars and fermions in $d=2$.

%Additional unproven inequalities involving the mutual information and the reflected entropy were suggested in  \cite{Dutta:2019gen}. The first reads
%\begin{equation}\label{xiii}
%\xi\equiv \frac{I(A,B)+I(A,C)}{R(A,BC)}\, , \quad \xi \overset{?}{\leq} 1\, ,
%\end{equation}
%whereas the second is given by
%\begin{equation}
%R(A_1 A_2,B_1B_2)\overset{?}{\geq} R(A_1,B_1)+R(A_2,B_2)\, .
%\end{equation}
%\comment{blah blahh}
%We have tested \req{xiii} in a few cases corresponding to $A$ being a fixed interval, $B$ corresponding to  

\section{Reflected entropy in $d=3$}\label{2plus1}
In this section we move to $(2+1)$-dimensional theories. In particular, we compute the reflected entropy for free massless scalars and fermions. We choose simple regions $A,B$ corresponding to parallel squares of length $L $  separated a distance $\ell$ along their bases. We study the behavior of $R(A,B)$ both for small and large values of $L/\ell$. For the latter, we extract the coefficients controlling the linear growth and compare them to the mutual information ones for both theories as well as for holographic Einstein gravity. Regarding the former, we observe a pattern, shared by the $d=2$ theories considered in the previous section, which leads us to conjecture that reflected entropy and mutual information for pairs of regions characterized by scales $L_A\sim L_B\sim L$ and separated a distance $\ell$ are universally related in the large-separation regime ($x\equiv L/\ell \ll 1$) by $R(x)\sim - I(x)\log x$ in general dimensions. 

%For the regions $A$,$B$ we choose two squares of side length $L$  separated a distance $\ell$. \comment{blah}

\subsection{Free scalar correlators}
In the case of the scalar, the Hamiltonian we have considered reads 
\begin{equation}
H=\frac{1}{2} \sum_{n,m=-\infty}^{\infty} \left[ \pi_{n,m}^2 + (\phi_{n+1,m}- \phi_{n,m})^2 + (\phi_{n,m+1} -\phi_{n,m})^2   \right]\, ,
\end{equation}
where the lattice spacing has been set to one. In this case, the formulation is in terms of bosonic fields and momenta, so the discussion in section \ref{tone} applies, and the relevant formulas for the reflected entropy are \req{phiii}, \req{entroo}. The relevant correlators read \cite{Casini:2009sr}
\begin{align}
X_{(0,0),(i,j)}\equiv \braket{\phi_{0,0}\phi_{i,j}} &=\frac{1}{8\pi^2} \int_{-\pi}^{\pi} \diff x \int_{-\pi}^{\pi} \diff y \frac{\cos (i x) \cos(j y)}{\sqrt{2(1-\cos(x))+2(1-\cos(y))}}\, , \\
P_{(0,0),(i,j)}\equiv\braket{\pi_{0,0}\pi_{i,j}} &=\frac{1}{8\pi^2} \int_{-\pi}^{\pi} \diff x \int_{-\pi}^{\pi} \diff y \cos(i x) \cos(jy) \sqrt{2(1-\cos x)+2(1-\cos y)}\, .
\end{align}
The subindices here refer to the coordinates of the corresponding two-dimensional lattice points. The correlators are invariant under translations, so that, $\braket{\phi_{0,0}\phi_{i,j}} =\braket{\phi_{n,m}\phi_{i+n,j+m}}$, and the same for the momenta.
For computational purposes, it is useful to perform the integral over $y$ in both expressions. The result can be written in terms of the regularized hypergeometric function ${}_p \tilde F_q$ as
\begin{align}
X_{(0,0),(i,j)} &=\frac{1}{2^{5/2}\pi } \int_{-\pi}^{\pi}  \diff x \frac{ \cos (i x) }{\sqrt{3-\cos x}} \, {}_3\tilde F_{2} \left[\{\tfrac{1}{2},\tfrac{1}{2},1\} ; \{1-j, 1+j\}; \frac{2}{3-\cos x} \right] \, , \\
P_{(0,0),(i,j)}&=\frac{1}{2^{3/2}\pi} \int_{-\pi}^{\pi} \diff x  \cos (i x) \sqrt{3-\cos x} \, {}_3\tilde F_{2} \left[\{\tfrac{1}{2},\tfrac{1}{2},1\} ; \{1-j, 1+j\}; \frac{2}{3-\cos x} \right]  \, .
\end{align}
These integrals can be easily evaluated numerically.

Regions  $A,B$ in the lattice correspond to subsets of points $p=(p_x,p_y)$. For instance, for a square region of length $L$ and with the lower left vertex at $(0,0)$, we have $A \equiv \{(p_x,p_y)\in \mathbb{Z}_2 \, |\, p_x,p_y=0,\dots, L \}$.
Given a pair of two-dimensional regions $A$ and $B$ we can evaluate the reflected entropy as follows. First, we need to evaluate the matrices $X$ and $P$. These are composed of four blocks corresponding to the $AA$, $AB$, $BA$ and $BB$ components, respectively. For instance, $X_{AB}$ corresponds to the block of eigenvalues $X_{p,q}$ where $p=(p_x, p_y)$, $q=(q_x,q_y)$  are points in the lattice such that  $p\in A$ and $q\in B$. Once we have $X$ and $P$, we need to evaluate $g(XP)$. In order to do that, we find the eigenvalues $\{ d_m\}$ of the matrix $XP$. Then, we build the diagonal matrix $\sqrt{d_m-1/4}\sqrt{d_m}^{-1}\delta_{mn}$ and transform it back to the original basis, which yields $g(XP)$. In order to obtain the off-diagonal blocks in $\Phi$ and $\Pi$, we multiply it by $X$ or $P$ as required in an obvious way. Finally, we restrict the matrices $\Phi$ and $\Pi$ to the $A$ region as
  \begin{align} \label{phiii0}
\Phi|_{AA^*}=\left(
\begin{array}{cc}
X|_A & \left[g(XP) X\right]|_A\\
\left[  g(XP) X\right]|_A & X|_A
\end{array}
\right)\, , \quad    \Pi|_{AA^*}=\left(
\begin{array}{cc}
P|_A & \left[-Pg(XP)\right]|_A  \\ 
\left[- Pg(XP) \right]|_A   & P|_A
\end{array}
\right)\, ,
\end{align}
where we used the notation $|_A$ to refer to the $AA$ block in each case. The final step is to evaluate $C_{AA^*}\equiv \sqrt{\Phi|_{AA^*} \Pi|_{AA^*}}$. Given the eigenvalues of this matrix, which we denote $\{\nu_m \}$, the reflected entropy finally reads 
\begin{equation}
R_{\rm scal.}= \sum_m  (\nu_m+1/2) \log (\nu_m+1/2)- (\nu_m-1/2)\log (\nu_m-1/2)\, .
\end{equation}

\subsection{Free fermion correlators}
For the $(2+1)$-dimensional Dirac fermion, the lattice Hamiltonian reads
\begin{equation}
H=- \frac{i}{2} \sum_{n,m}  \left[\left( \psi_{m,n}^{\dagger} \gamma^0 \gamma^1 (\psi_{m+1,n} - \psi_{m,n}) + \psi^{\dagger}_{m,n}\gamma^0\gamma^2 (\psi_{m,n+1}-\psi_{m,n}) \right)- h.c. \right]\, ,
\end{equation}
and the corresponding correlators \cite{Casini:2009sr}
\begin{equation}
\braket{\psi^{\dagger}_{n,k} \psi_{j,l}}=\frac{1}{2} \delta_{nj}\delta_{kl} + \int_{-\pi}^{\pi} \diff x \int_{-\pi}^{\pi} \diff y \frac{\sin( x) \gamma^0 \gamma^1 +\sin (y) \gamma^0\gamma^2}{8\pi^2\sqrt{\sin^2x+\sin^2y}}e^{i (x (n-j)+ y (k-l))}\, .
 \end{equation}
 Just like in the case of the scalar, the subindices in the fermionic fields above correspond to the coordinates of the corresponding lattice points.
 
The relevant formulas for the evaluation of the reflected entropy in the case of fermionic Gaussian systems were obtained in \cite{Bueno:2020vnx}. Let us quickly summarize the relevant results here. We start with $N$ fermions, $\psi_i$, $i=1,\dots,N$, satisfying canonical anticommutation relations $\{\psi_i,\psi^{\dagger}_j \}=\delta_{i,j}$ and a density matrix $\rho$ in the corresponding Hilbert space of dimension $2^N$. We can purify this state by doubling the Hilbert space and use the modular reflection operator $J$ associated to such state and the algebra of the first $N$ fermions to double the fermion algebra ---doing this properly involves a unitary constructed from the fermion number operator \cite{Doplicher:1969tk}--- in a way such that we are left with a canonical set of $2N$ operators. Denoting by $D_{ij}\equiv \tr (\rho \psi_i \psi_j^{\dagger})$ the correlators of the original system, the matrix which turns out to be relevant for the evaluation of reflected entropy is given by 
\begin{equation}
C=\left(
\begin{array}{cc}
 D & \sqrt{D(1-D)} \\
\sqrt{D(1-D)} &  (1- D
\end{array}
\right)\, ,\label{cij}
\end{equation}
where the additional blocks correspond to the appearance of new correlators involving the new fermionic fields in the doubled system. Just like in the case of the scalars, the final answer can be fully written in terms of correlators of the original system, as is apparent in \req{cij}. Finally, the reflected entropy for a pair of regions $A$,$B$ is obtained from the restrictions of the corresponding block matrices to $A$ 
%Type-I algebra $\mathcal{N}\equiv \mathcal{A}_A \vee J_{AB} \mathcal{A}_A J_{AB}$ where $D$ is replaced by $C_{AA*}$ \cite{Bueno:2020vnx} 
\begin{equation}
C_{AA^*}=\left(
\begin{array}{cc}
 D|_A& \left.\sqrt{D(1-D)}\right|_A\\
\left.\sqrt{D(1-D)}\right|_A &  (1- D)|_A
\end{array}
\right)\, .\label{cija}
\end{equation}
Denoting by $\{ \nu_m \}$ the eigenvalues of $C_{AA^*}$, we finally have
\begin{equation}
R_{\rm ferm.}=- \sum_m \left[ \nu_m \log(\nu_m)+ (1- \nu_m)\log(1-\nu_m)\right]\, .
\end{equation}
When taking the continuum limit, we have to take into account the doubling of the fermionic degrees of freedom on the lattice. In $(2+1)$ dimensions, this requires dividing the final result by $4$ in order to obtain the result corresponding to a Dirac fermion. When presenting the results, we will consider reflected entropy (or mutual information) per degree of freedom, which in this case requires dividing by an addition factor of $2$.  %\comment{blah blahhh} 

\subsection{Reflected entropy for two parallel squares}
Using the results of the previous two subsections, we are ready to evaluate the reflected entropy of scalar and fermionic systems in $(2+1)$ dimensions. We do this for regions $A$,$B$ corresponding to two squares of length $L$ aligned so that the second square can be obtained by moving the first a distance $L+\ell$ along the (positive) direction of its base. We have then two parallel squares separated by a distance $\ell$. The corresponding sets in the lattice correspond to  $A \equiv \{(p_x,p_y)\in \mathbb{Z}_2 \, |\, p_x,p_y=0,\dots, L \}$ and  $B \equiv \{(p_x,p_y)\in \mathbb{Z}_2 \, |\, p_x=L+\ell,\dots, 2L+\ell\, , \, p_y=0,\dots,L \}$.

\begin{figure}[t] \centering
	\includegraphics[scale=0.65]{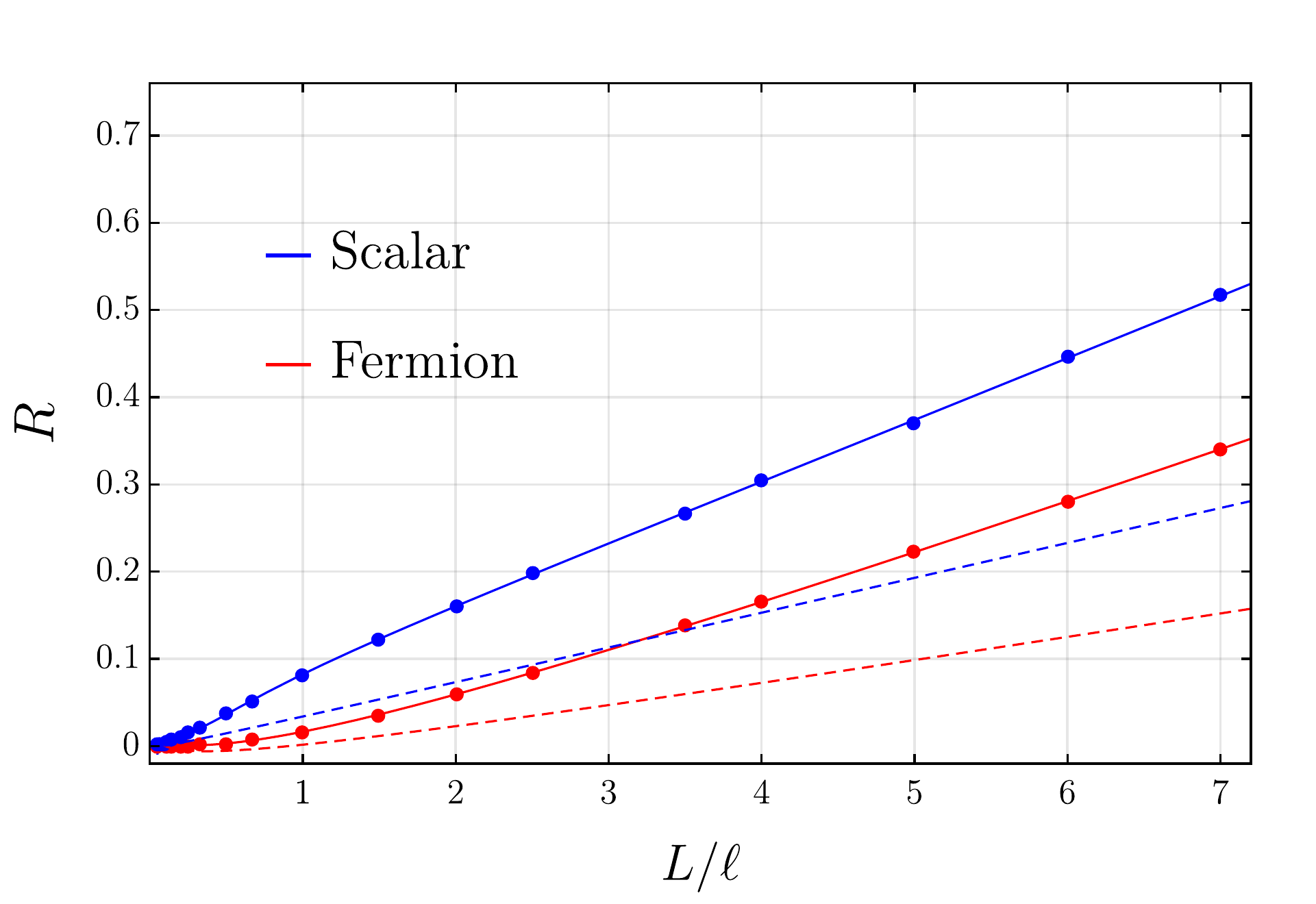}
	\caption{We plot the reflected entropy (per degree of freedom) for regions $A$, $B$, corresponding to two squares of length $L$ separated by a distance $\ell$ as a function of $L/\ell$ for a free scalar (blue) and a free fermion (red). For both fields we also plot the mutual information $I(A,B)$ for the same pair of regions (dashed lines). The latter curves are obtained numerically using the usual definition $I(A,B)=\see(A)+\see(B)-\see(AB)$, where the corresponding entanglement entropies are computed in the lattice using the same von Neumann entropy formulas as for the reflected entropies, but associated to $\rho_A$ instead of $\rho_{AA^*}$ in each case.   }
	\label{refiss331}
\end{figure}

Using the procedures explained in the previous two subsections, we obtain the results shown in Fig.\,\ref{refiss331} for the corresponding reflected entropies as a function of the quotient $L/\ell$. Just like it happens for the mutual information ---also shown in the plots--- the scalar result is greater than the fermion one in the whole range of values. Also, in both cases, we find that the general inequality \req{ir} holds. In the case of the scalar, it is actually possible to obtain reflected entropy using the formulas in subsection \ref{take2} instead of those in subsection \ref{tone}. We have done so and verified that the results agree, which is a good consistency check for our general formulas. 

\begin{figure}[t] \centering
\includegraphics[scale=0.642]{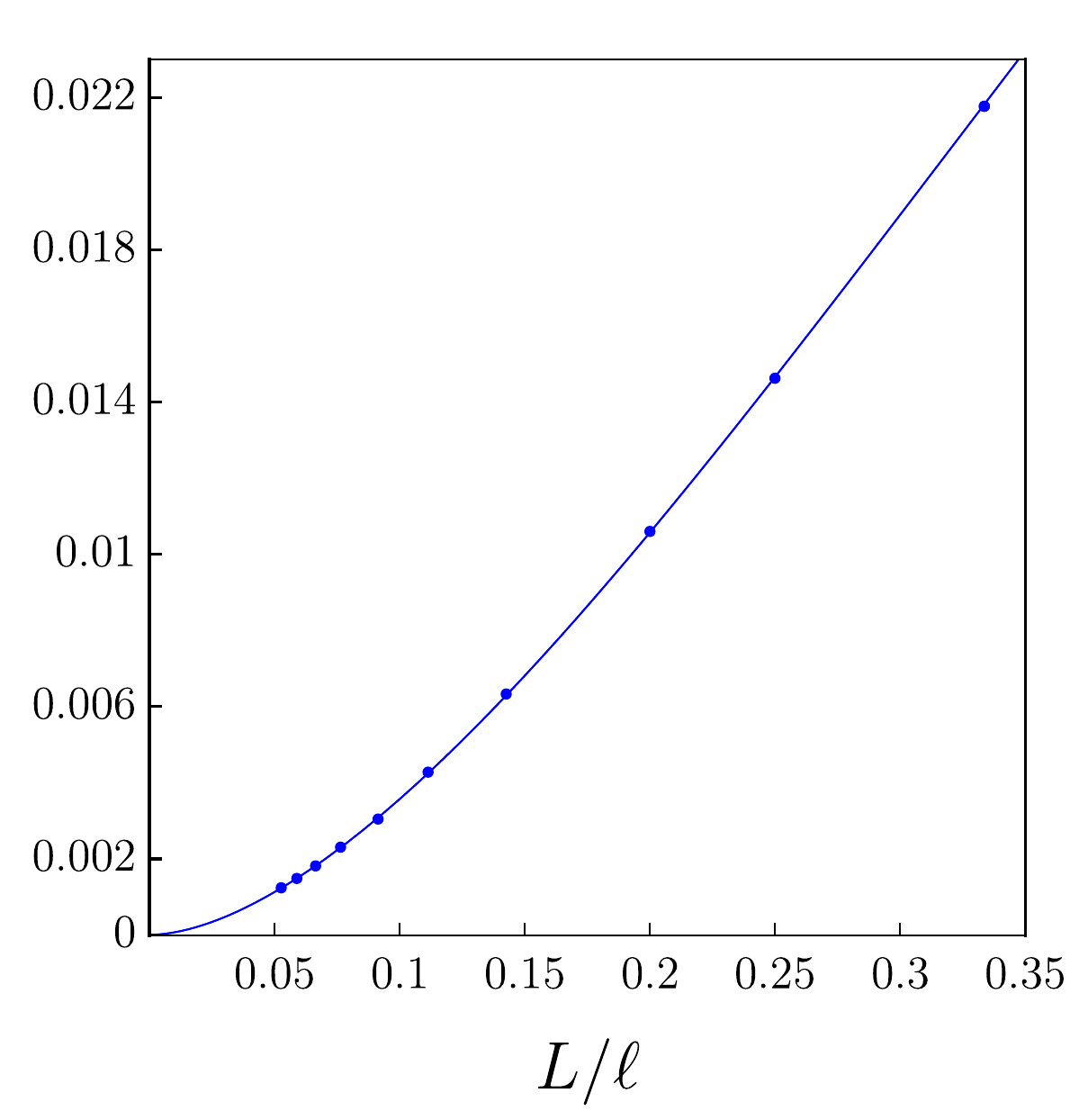}
	\includegraphics[scale=0.645]{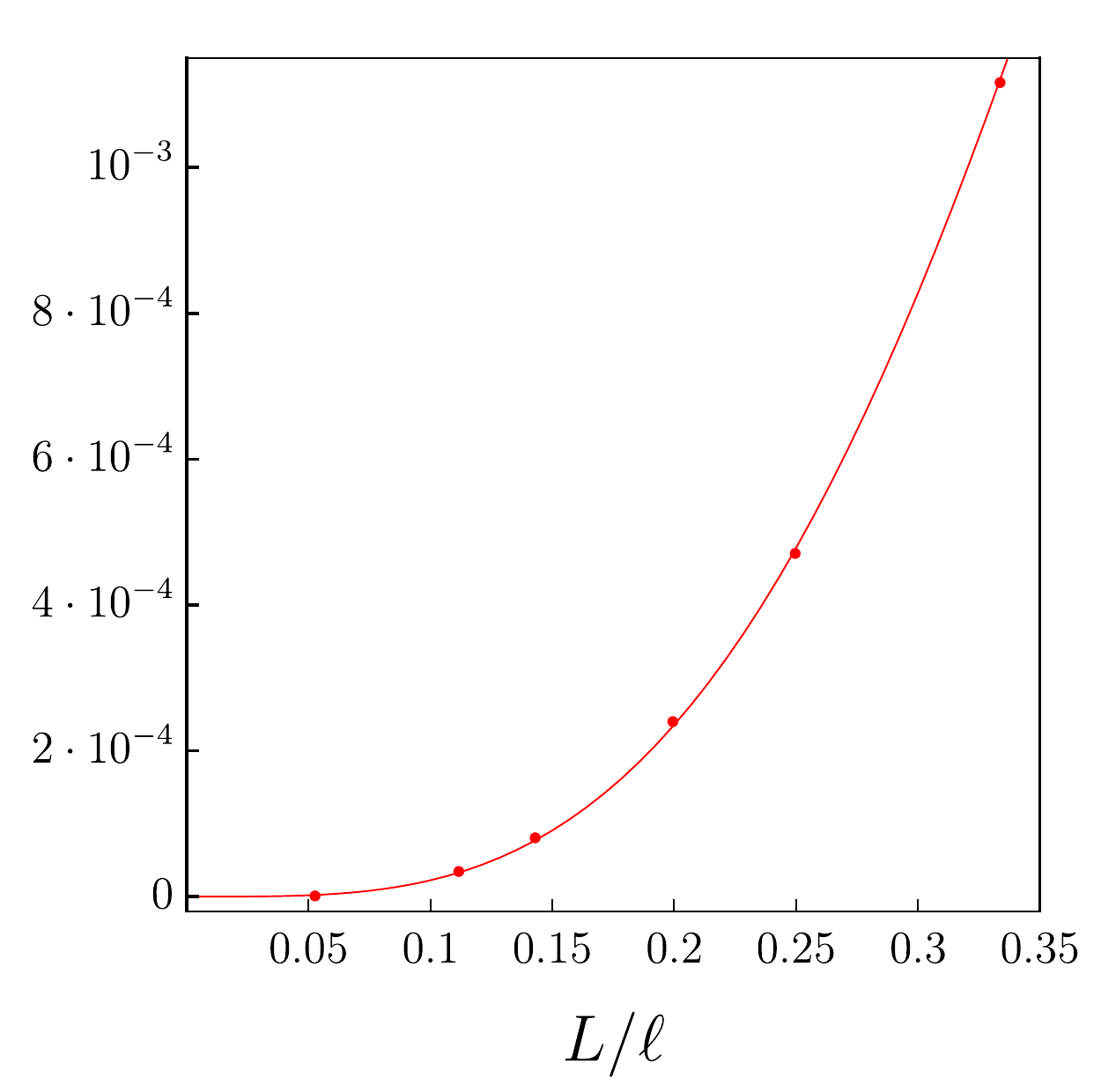}
	\caption{We plot the reflected entropy (per degree of freedom) for regions $A$, $B$, corresponding to two squares of length $L$ separated by a distance $\ell$ as a function of $L/\ell$ for a free scalar (blue dots) and a free fermion (red dots) in the small-$L/\ell$ region. We also show the trial functions explained in the text.}
	\label{refiss33d1}
\end{figure}

For small values of $x\equiv L/\ell$ we do not have a priori a clear guess of what the behavior of $R_{\rm scal.}$ and $R_{\rm ferm.}$ should be. We have looked for trial functions involving simple combinations of powers and logarithms and such that: they go to zero at $x=0$, they are positive in the whole range, they grow monotonically in the domain considered, the fit coefficients are neither too large nor too small. In the case of the scalar, we find that the following function does a good job in fitting the numerical data 
%\comment{near zero behavior?} 
\begin{align}\label{sepas}
R_{\rm scal.}(x \ll 1) &\sim - 0.133 x^2 \log x+ 0.0497 x^2 \, .
%R_{\rm ferm.} &\simeq
\end{align}
We plot this function alongside the numerical data points in Fig.\,\ref{refiss33d1}. As we can see, the fit is actually good up to values $x\lesssim 0.38$.  In the case of the fermion, we find that the following fit approximates well the data points up to similar values of  $x$
\begin{align}\label{sepaf}
R_{\rm ferm.}(x \ll 1) &\sim - 0.111 x^4 \log x-0.03144 x^4  \, .
%R_{\rm ferm.} &\simeq
\end{align}
This appears shown in the right plot in Fig.\,\ref{refiss331}. It is interesting to compare these expressions with the corresponding mutual information behavior. For that, we note that given two regions with characteristic scale $L$ separated by a much larger distance $\ell$, one finds for general $d$-dimensional CFTs  \cite{Calabrese:2009ez,Calabrese:2010he,Cardy:2013nua,Agon:2015ftl}
\begin{equation}\label{ix}
I (x \ll 1)\sim x^{4\Delta}\, ,
\end{equation}
where $\Delta$ is the scaling dimension of the lowest-dimensional operator of the  corresponding theory. Hence, for scalars and fermions we have
\begin{equation}
I_{\rm ferm.} (x \ll 1)\sim x^{2(d-1)}\, , \quad I_{\rm scal.} (x \ll 1)\sim x^{2(d-2)}\, ,
\end{equation}
respectively. Thus, we observe that $I_{\rm ferm.} (x \ll 1)\sim x^4$ and $I_{\rm scal.} (x \ll 1)\sim x^{2}$ in the three-dimensional case considered here. Comparing with  \req{sepas} and \req{sepaf}, we observe that reflected entropy behaves with the same power as mutual information  multiplied by a logarithm of $L/\ell$. Going back to section \ref{chiralsca}, we observe that exactly the same phenomenon is found both for the chiral scalar\footnote{Note that in the case of the chiral scalar considered in section \ref{chiralsca}, the lowest-dimensional operator is $\partial \phi$, for which $\Delta=1$.} and the fermion. These results are very suggestive and lead us to propose the following conjecture.

{\bf Conjecture:} The reflected entropy for two regions $A,B$ with characteristic scales $L_A\sim L_B\sim L$ separated a distance $\ell$ behaves as
\begin{equation}\label{conje}
R(x) \sim -I(x) \log x\sim  - x^{4\Delta} \log x\, ,  \quad (x\equiv L/\ell)\, ,
\end{equation}
in the $x \ll 1$ regime for general CFTs in arbitrary dimensions.

It would be interesting to test the validity of this conjectural relation for additional models in various dimensions (as well as for higher-dimensional free-field theories) or to (dis)prove it in general. A natural setup where \req{conje} could be tested would be holography. In that case, the leading-order result of both reflected entropy and mutual information vanishes for sufficiently small values of $x$ (\eg for $\eta< 1/2$ in the intervals case in $d=2$). Accessing the first non-vanishing contribution in the mutual information case in that regime requires considering quantum corrections to the Ryu-Takayanagi formula \cite{Faulkner:2013ana}, and the result agrees with the general CFT behavior in \req{ix} \cite{Agon:2015ftl}. An analogous expression for the leading correction of holographic reflected entropy was presented in \cite{Dutta:2019gen}, so it should be in principle possible to check the validity of our conjecture in that case.

For large values of $L/\ell$, the behavior of $R(A,B)$  becomes linear. The reason for this is that, as the length of the squares grows with respect to the separation, the setup becomes more and more similar to the case of two infinitely-extended parallel sets for which the leading contribution is an ``area-law'' like term. The situation is analogous for the mutual information, and the corresponding linear growth is also apparent in the corresponding dashed lines in Fig. \ref{refiss331}. More generally, for any $d$-dimensional CFT, when $A,B$ are two sets with large parallel faces of area $\mathcal{A}$ separated by a comparatively small distance $\ell$ one finds
%\subsection{Two sets with large parallel faces}
%When $A,B$ are two sets with large parallel faces of area $\mathcal{A}$ separated by a comparatively small distance $\ell$, the mutual information behaves as
\begin{equation}
I(A,B)= \kappa^{(I)}_d \frac{\mathcal{A}}{\ell^{d-2}}+ \text{subleading}\, , \quad R(A,B) = \kappa^{(R)}_d \frac{\mathcal{A}}{\ell^{d-2}}+ \text{subleading}\, .
\end{equation}
As shown in \cite{Casini:2009sr,Casini:2005zv}, both for free scalars and fermions, the values for the mutual information coefficients $\kappa^{(I)}_d$ can be obtained from a dimensional reduction to $(1+1)$ dimensions. The results are given in terms of the functions appearing in the entropic version of the $c$-theorem\footnote{This is defined from the entanglement entropy of an interval of length $L$ as $c(L)\equiv L \tfrac{d\see (L)}{dL}$.}  \cite{Casini:2004bw} corresponding to the respective free theories in that number of dimensions. The explicit results in $d=3,4,5,6$ for both types of fields read
\begin{align}
\kappa^{(I)}_{3, {\rm \, sc.}}\simeq3.97\cdot 10^{-2} \, , \quad \kappa^{(I)}_{4, {\rm \, sc.}}\simeq 5.54 \cdot 10^{-3} \, , \quad \kappa^{(I)}_{5, {\rm \, sc.}}\simeq1.31\cdot 10^{-3} \, , \quad \kappa^{(I)}_{6, {\rm \, sc.}}\simeq4.08 \cdot 10^{-4} \, ,  \\ 
\kappa^{(I)}_{3, {\rm \, fer.}}\simeq3.61\cdot 10^{-2}  \, , \quad \kappa^{(I)}_{4, {\rm \, fer.}}\simeq5.38 \cdot 10^{-3} \, , \quad \kappa^{(I)}_{5, {\rm \, fer.}}\simeq1.30\cdot 10^{-3} \, , \quad \kappa^{(I)}_{6, {\rm \, fer.}}\simeq4.06 \cdot 10^{-4}  \, .
\end{align}
As $d\rightarrow \infty$, the scalar and fermion results tend to a common value, given by \cite{Casini:2009sr}
\begin{equation}
\kappa^{(I)}_{d \rightarrow \infty}=\frac{\Gamma\left[\tfrac{d-2}{2} \right]}{2^{d+2} \pi^{\tfrac{d-2}{2}}}\, .
\end{equation}
Naturally, the $d=3$ coefficients are the slopes of the leading contributions to the dashed curves shown in Fig. \ref{refiss331} as $L/\ell \gg 1$. In order to extract these values from the numerical results, we perform a fit with a linear, a logarithmic and a constant function to the data points obtained with $L/\ell >4$. The results obtained from numerical fits are in good agreement with the values of $\kappa^{(I)}_{3, {\rm \, scal.}}$ and $\kappa^{(I)}_{3, {\rm \, ferm.}}$ shown above. Proceeding similarly for the reflected entropy, we find
\begin{equation}
\kappa^{(R)}_{3, {\rm \, scal.}}\simeq 6.95\cdot 10^{-2} \, , \quad \kappa^{(R)}_{3, {\rm \, ferm.}}\simeq 6.16\cdot 10^{-2}\, .
\end{equation}
%\comment{reflected  result, quotient}

We can compare these results to holographic theories dual to Einstein gravity, for which the values of $\kappa^{(R)}_{d, {\rm \, holo.}}$ and $\kappa^{(I)}_{d, {\rm \, holo.}}$ can be obtained analytically in general dimensions. In the case of the mutual information, the coefficient can be extracted from the 
%On the other hand, the holographic result for theories dual to Einstein gravity can be obtained analytically for general values of $d$ from the 
universal term in the entanglement entropy corresponding to a strip of width $\ell$ much smaller than the rest of dimensions. The bulk action reads
\begin{equation}
I_g=\frac{1}{16\pi G} \int d^{d+1}x \sqrt{|g|} \left[\frac{d(d-1)}{L^2}+R \right]\, ,
\end{equation}
where $G$ is the Newton constant and where we parametrized the cosmological constant so that  AdS$_{(d+1)}$ is a solution of the theory with radius $L$. Entanglement entropy can be then obtained from the Ryu-Takayanagi prescription \cite{Ryu:2006bv,Ryu:2006ef} and the relevant coefficient turns out be given by \cite{Ryu:2006bv}
%, and is given by
\begin{equation}
\kappa^{(I)}_{d, {\rm \, holo.}}= \frac{2^{d-3}\pi^{\tfrac{d-1}{2}} \Gamma\left[\tfrac{d}{2(d-1)} \right]^{d-1} }{(d-2) \Gamma\left[\tfrac{1}{2(d-1)} \right]^{d-1}} \frac{L^{d-1}}{G}\, .
\end{equation}
%where $L_{\star}$ is the AdS$_{(d+1)}$ radius and $G$ is Newton's constant.
%Now, just like mutual information, reflected entropy can be used as a regulator for entanglement entropy \cite{Dutta:2019gen}. As a consequence, the same kind of behavior for two regions with large parallel faces is encountered. Namely, one has
%Now, the reflected entropy also satisfies the same kind of behavior when $A$, $B$ are sets of the kind described above. One has 
%\begin{equation}
%R(A,B) = \kappa^{(R)}_d \frac{\mathcal{A}}{\ell^{d-2}}+ \text{subleading}\, .
%\end{equation}
%The coefficients $\kappa^{(R)}_d$ will in general differ from their mutual information counterparts. In the case of holographic theories dual to Einstein gravity, 
In the case of the reflected entropy, we can obtain the result assuming its relation to the minimal entanglement wedge cross section, $R_{\rm holo.}(A,B)=2E_W(A,B)$ proposed in \cite{Dutta:2019gen}. This calculation was performed in  \cite{Jokela:2019ebz} in the more general case of two parallel strips of fixed width. Taking the large-width limit of the result we can extract $\kappa^{(R)}_{d, {\rm \, holo.}}$. The result reads
\begin{equation}
\kappa^{(R)}_{d, {\rm \, holo.}}= \frac{2^{d-3} \pi^{\tfrac{d-2}{2}}\Gamma\left[\tfrac{d}{2(d-1)} \right]^{d-2}  }{(d-2) \Gamma\left[\tfrac{1}{2(d-1)} \right]^{d-2}} \frac{L^{d-1}}{G}\, .
\end{equation}
In order to compare with the free-field results, we can consider the quotient between both coefficients, which reads
%Then, we find for the quotient
\begin{equation}
\frac{\kappa^{(R)}_{d, {\rm \, holo.}}}{\kappa^{(I)}_{d, {\rm \, holo.}}}= \frac{\Gamma\left[\tfrac{1}{2(d-1)} \right]}{\sqrt{\pi} \Gamma\left[\tfrac{d}{2(d-1)} \right]}\, .
\end{equation}
This is always larger than $1$, as it should in view of the inequality \req{ir}. In particular, 
\begin{equation}
\frac{\kappa^{(R)}_{3, {\rm \, holo.}}}{\kappa^{(I)}_{3, {\rm \, holo.}}}\simeq 1.669 \, , \quad \frac{\kappa^{(R)}_{4, {\rm \, holo.}}}{\kappa^{(I)}_{4, {\rm \, holo.}}}\simeq 2.319 \, , \quad \frac{\kappa^{(R)}_{5, {\rm \, holo.}}}{\kappa^{(I)}_{5, {\rm \, holo.}}}\simeq 2.963\, , \quad  \frac{\kappa^{(R)}_{6, {\rm \, holo.}}}{\kappa^{(I)}_{6, {\rm \, holo.}}}\simeq 3.604\, .
\end{equation}
As $d\rightarrow \infty$, one has 
\begin{equation}
\frac{\kappa^{(R)}_{d\rightarrow \infty, {\rm \, holo.}}}{\kappa^{(I)}_{d\rightarrow \infty, {\rm \, holo.}}}= \frac{1}{\pi} \left[2d+(\log 4-2) + \mathcal{O} (1/d) \right]\, .
\end{equation}
%Using the numerical calculations 
The $d=3$ result is not so different from the ones we find numerically for the free fields. For those, we obtain
%Our numerical calculations for free scalars and fermions yield
\begin{equation}
\frac{\kappa^{(R)}_{3, {\rm \, scal.}}}{\kappa^{(I)}_{3, {\rm \, scal.}}}\simeq 1.75  \, , \quad \quad \frac{\kappa^{(R)}_{3, {\rm \, ferm.}}}{\kappa^{(I)}_{3, {\rm \, ferm.}}}\simeq 1.71 \, .
\end{equation}
Some degree of similarity between the free fermion and holography ---as far as entropic measures are concerned--- has been previously observed in other situations ---see \eg \cite{Bueno:2015qya}. Here we observe that the fermion result is indeed more similar to the holographic answer, but it is not extremely close to either of the two.

%\comment{comparisooooonnn comments} 

While the leading contribution for large values of $L/\ell$ is linear, there is a subleading logarithmic term. The presence of logarithmic contributions associated to corner regions is characteristic of this kind of measures. In the  entanglement entropy case, the corresponding universal contribution has been subject of intense study ---see \eg \cite{Bueno:2019mex} for an updated list of relevant references. While we have not attempted to evaluate with reasonable numerical precision the logarithmic terms appearing in the case of the two square regions considered here for the reflected entropy, we point out that we do not expect the corresponding pieces to be immediately related to entanglement entropy corner terms corresponding to a single square region (as opposed to mutual information, for which they are). In order to extract such term from a reflected entropy calculation, we would need to consider regions $A,B$ corresponding to a square and the complement of a larger square, respectively. On the other hand, this also means that reflected entropy contains new universal coefficients with no immediate entanglement entropy counterpart.
We leave a study of such kind of terms for future work.

\section{Outlook}\label{finnnn}
As we have illustrated, the formulas presented here and in \cite{Bueno:2020vnx} allow for simple numerical evaluations of reflected entropy for free scalars and fermions. While the expressions are valid in general dimensions, our analysis so far has been mostly focused  on two-dimensional theories. In section \ref{2plus1} we made a first incursion into higher dimensions, but we restricted ourselves to parallel square-like regions. It would be interesting to continue exploring higher-dimensional theories and the various universal terms appearing associated to different kinds of regions ---\eg the coefficients $\kappa^{(R)}_d$ for $d>3$. As mentioned above, this will include terms with an without entanglement entropy counterparts.

Another direction would entail considering massive theories. In most cases, the relevant correlators are simple (and known) modifications of the ones we have used here for the corresponding massless cases, so such generalizations are clearly accessible. 
 % There are various directions for future research that we would like to mention before closing.

It would also be interesting to better clarify the  connection and differences between reflected entropy and other entanglement measures. In particular, it would be nice to test the validity of our conjectural relation \req{conje} for additional theories. In fact, perhaps a general proof could be attempted using Replica-trick methods \cite{Dutta:2019gen}.  Beyond mutual information, connections between reflected entropy and odd-entropy have also been reported \cite{Tamaoka:2018ned,Berthiere:2020ihq,Mollabashi:2020ifv}, which would be interesting to examine further. 

Finally, in \cite{Bueno:2020vnx} we introduced a modification of reflected entropy ---``type-I entropy''--- which differed from the former in the case of theories obtained from quotients of theories  by global symmetry groups. Operators implementing the corresponding symmetry operations on type-I algebras can be constructed and computationally amenable notions of entropy can be associated to their expectation values. Those connect in a simple fashion the reflected entropies of complete theories and the type-I entropies of subalgebras. This suggests possible interesting entropic connections between bosonic and fermionic theories related by quotients. 

We plan to explore some of these directions in the near future.

\section*{Acknowledgments}
We thank Cl\'ement Berthiere, Robie A. Hennigar and Javier M. Mag\'an for useful discussions. 
This work was supported by the Simons foundation through the It From Qubit Simons collaboration.
H.C. was also supported by CONICET, CNEA, and Universidad Nacional de Cuyo, Argentina.

\appendix
%\section{Purification of free fermions}
%Summary of purification of free fermions from other paper

\section{Numerical values of $R_{\rm ferm.}/c$ and $R_{\rm scal.}/c$ }\label{ape}
In this appendix we present numerical results found for the reflected entropy of two intervals $A$,$B$ for different values of the cross-ratio ---defined in \req{cr7}--- in the case of a free fermion and a free scalar in $1+1$ dimensions. The values presented here are those shown in Fig.\ref{refiss1}. Results are presented from smaller to greater values of $\eta$. For technical reasons, in some cases we chose slightly different values of $\eta$ to evaluate the reflected entropy for each field. Also, as $\eta$ approaches one, obtaining reliable numerical values becomes increasingly demanding, which is why we present less significant digits in that case. 
\begin{center} \footnotesize	\begin{tabular}{|c|c|c|}
	\hline % changed
	$\eta$	&  $R_{\rm ferm.}/c$ 
		&   $R_{\rm scal.}/c$  % changed
		    \\ \hline \hline
	0   &   0 &  0
		\\ \hline
		$1/121$   &  0.01146 &   0.00001572
		\\ \hline
		$1/100$   &   0.01359 &   0.00002249
		\\ \hline
		$1/49$   &   0.02569 &  0.00008579
		\\ \hline
		$1/9$   &   0.11603 &   0.002075
		\\ \hline
		$4/25$   &   0.16150 &  0.004177
		\\ \hline
		$1/4$   &   0.2453 &   0.01005
		\\ \hline
		$625/1936$   &    &  0.01696
		\\ \hline
		$16/49$   &  0.3167  & 
		\\ \hline
		$4/9$   &   0.4347 &  0.03391
		\\ \hline
		$10000/18769$    &   0.5315 &  0.0518
		\\ \hline
		$16/25$   &   0.6648 &  0.0833
		\\ \hline
		$25/36$   &    &  0.105
		\\ \hline
		$10000/13689$   & 0.7999    &  
		\\ \hline
		$625/841$   &    &  0.130
		\\ \hline
		$64/81$   &  0.9077   &  
		\\ \hline
		$625/784$   &    &  0.166
		\\ \hline
		$100/121$   &    &  0.191
		\\ \hline
		$10000/11881$  &  1.021   &  
		\\ \hline
		$625/729$   &    &  0.223
		\\ \hline
		$10000/11449$  &  1.108   &  
		\\ \hline
		$2500/2809$   &    &  0.268
		\\ \hline
		$400/441$   & 1.222   &  
		\\ \hline
		$625/676$   &   1.299 &  0.335
		\\ \hline
		$10000/10609$   &  1.396 &  0.382
		\\ \hline
		$2500/2601$   &   1.54 &  0.462
		\\ \hline
		$10000/10201$   &    &  0.630
		\\ \hline
		$40000/40401$   &    &  0.78
		\\ \hline
	\end{tabular}
	\label{bhtaub}
\end{center}

\bibliographystyle{JHEP}
\bibliography{Gravities}
%\label{biblio}

\end{document}